\documentclass[pra,showpacs,twocolumn,aps,superscriptaddress,a4paper]{revtex4-1}
\usepackage{dcolumn,amssymb,amsmath,amsfonts,graphicx,latexsym,color}
\usepackage{epstopdf}
\usepackage{CJK}
\begin{document}

\begin{CJK*}{GBK}{song}

\title{Three-component Fulde-Ferrell superfluids in a two-dimensional Fermi gas with spin-orbit coupling}

\author{Fang Qin}
\email{qinfang@phy.ccnu.edu.cn}
\affiliation{Key Laboratory of Quantum Information, University of Science and Technology of China, Chinese Academy of Sciences, Hefei, Anhui 230026, China}
\affiliation{School of Mathematics and Physics and Institute for Quantum Materials, Hubei Polytechnic University, Huangshi, Hubei 435003, China}

\author{Fan Wu}
\affiliation{Key Laboratory of Quantum Information, University of Science and Technology of China, Chinese Academy of Sciences, Hefei, Anhui 230026, China}
\affiliation{Synergetic Innovation Center of Quantum Information and Quantum Physics, University of Science and Technology of China, Hefei, Anhui 230026, China}

\author{Wei Zhang}
\email{wzhangl@ruc.edu.cn}
\affiliation{Department of Physics, Renmin University of China, Beijing 100872, China}
\affiliation{Beijing Key Laboratory of Opto-electronic Functional Materials and Micro-nano Devices,
Renmin University of China, Beijing 100872, China}

\author{Wei Yi}
\email{wyiz@ustc.edu.cn}
\affiliation{Key Laboratory of Quantum Information, University of Science and Technology of China, Chinese Academy of Sciences, Hefei, Anhui 230026, China}
\affiliation{Synergetic Innovation Center of Quantum Information and Quantum Physics, University of Science and Technology of China, Hefei, Anhui 230026, China}

\author{Guang-Can Guo}
\affiliation{Key Laboratory of Quantum Information, University of Science and Technology of China, Chinese Academy of Sciences, Hefei, Anhui 230026, China}
\affiliation{Synergetic Innovation Center of Quantum Information and Quantum Physics, University of Science and Technology of China, Hefei, Anhui 230026, China}
\date{\today}

\begin{abstract}
We investigate the pairing physics of a three-component spin-orbit coupled Fermi gas in two spatial dimensions. The three atomic hyperfine states of the system are coupled by the recently realized synthetic spin-orbit coupling (SOC), which mixes different hyperfine states into helicity branches in a momentum-dependent manner. As a consequence, the interplay of spin-orbit coupling and the hyperfine-state dependent interactions leads to the emergence of Fulde-Ferrell (FF) pairing states with finite center-of-mass momenta even in the absence of the Fermi-surface asymmetry that is usually mandatory to stabilize an SOC-induced FF state. We show that, for different combinations of spin-dependent interactions, the ground state of the system can either be the conventional Bardeen-Cooper-Schrieffer pairing state with zero center-of-mass momentum or be the FF pairing states. Of particular interest here is the existence of a three-component FF pairing state in which every two out of the three components form FF pairing. We map out the phase diagram of the system and characterize the properties of the three-component FF state, such as the order parameters, the gapless contours and the momentum distributions. Based on these results, we discuss possible experimental detection schemes for the interesting pairing states in the system.
\end{abstract}


\maketitle

\section{Introduction}

Ever since its experimental realization, synthetic spin-orbit coupling (SOC) in ultracold atomic gases has attracted much attention ~\cite{gauge2exp,shuaiexp1,engels,shuaiexp2,fermisocexp1,fermisocexp2,fermisocexp3}. Over the past few years, SOC-induced exotic phases and phase transitions have been extensively studied in both the Bose and the Fermi gases ~\cite{zhaireview2012,galitskireview,congjunreview,spielmanreview,zhaireview2015,hupureview,chuanweireview,wyreview,Mannarellireview}. Central to the effects of SOC in these systems is the nonperturbative modification of the single-particle dispersion spectra. For Fermi gases in particular, the SOC-modified single-particle dispersion can induce exotic few-body states~\cite{cuizhaiprl,cuiprx} as well as highly nontrivial many-body correlations~\cite{wu20131,yi2014PRL}. The SOC-induced Fulde-Ferrell (FF) pairing state is an interesting example where the interplay of SOC and the Zeeman-field-induced Fermi surface asymmetry stabilizes unconventional pairing superfluids with finite center-of-mass momenta~\cite{wyreview}.

Recently, it has been shown that an alternative FF pairing mechanism exists in a Fermi-Fermi mixture where a two-component noninteracting Fermi gas, while dressed by SOC, interacts spin selectively with a third fermionic species~\cite{yi2014PRL}. As the hyperfine-spin distribution in the SOC-induced helicity branches is asymmetric in momentum space, the spin-selective pairing interaction leads to pairing states with finite center-of-mass momenta in both the two-body and the many-body sectors. This novel pairing mechanism is thus fundamentally different from the majority of existing proposals of SOC-induced FF states since it is originated from spin-selective interaction and does not require a Fermi-surface asymmetry.

In this paper, we extend this exotic spin-selective-interaction-induced FF mechanism to a three-component Fermi gas where all three hyperfine states are coupled by SOC~\cite{lan,gao2014}. From the single-particle dispersion, we show that in the three SOC-induced helicity branches, two exhibit asymmetric hyperfine-spin distributions. Based on this asymmetry, we discuss several different configurations of the spin-selective interactions, under which the ground state of the system can either be a conventional Bardeen-Cooper-Schrieffer (BCS) state with zero center-of-mass momentum or be FF states originating from the interplay of SOC and pairing interaction. In particular, an interesting three-component FF state can be identified in which every two out of the three hyperfine components form FF pairing with a common center-of-mass momentum.  We investigate the stability and phase transitions of the three-component FF states by mapping out the mean-field phase diagram. Depending on the excitation gap of quasiparticles, the FF states in the present system can be further categorized as gapless or fully gapped. We characterize the gapless FF states by studying the gapless contours in momentum space and discuss their relation with features in the number distribution, which should facilitate experimental detection. As the various configurations of interaction can in principle be experimentally implemented via the Feshbach resonance technique, our paper not only reveals the generality of the FF pairing mechanism induced by SOC and spin-selective interactions, but also has interesting implications for future experiments on SOC-induced exotic superfluidity.

The paper is organized as follows: In Sec.~\ref{2}, we calculate the single-particle dispersion under SOC and analyze the inherent asymmetry in the
hyperfine-state distribution. In Sec.~\ref{3}, we present the system as well as the mean-field formalism. We then consider different configurations of spin-dependent interactions and investigate their impact on the pairing superfluidity in Sec.~\ref{4}. In Sec.~\ref{5}, we discuss the novel three-component FF state. The main results of this paper are summarized in Sec.~\ref{6}.

\section{Single-particle dispersion}\label{2}

\begin{figure}[tbp]
\includegraphics[width=6cm]{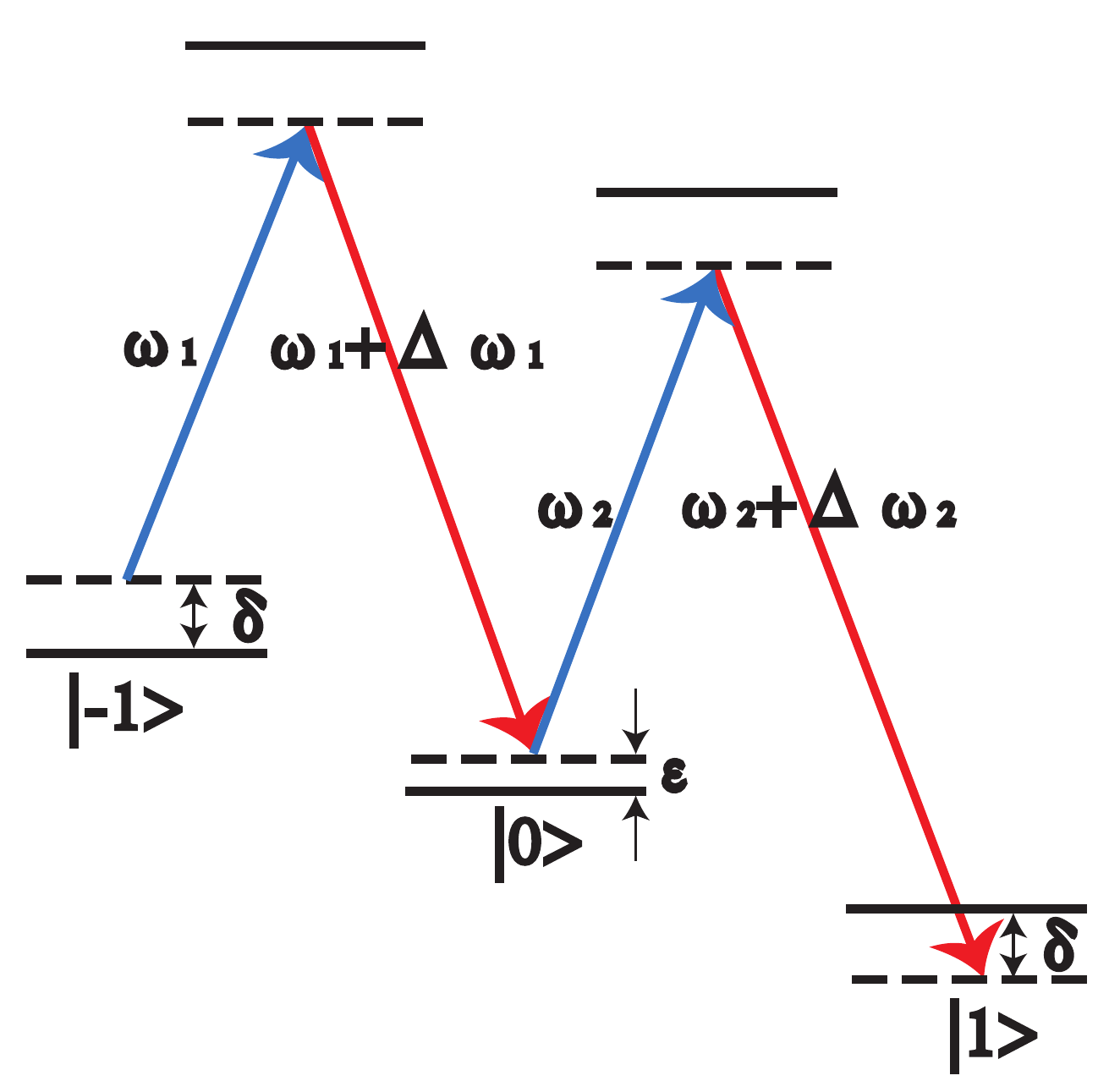}
\caption{(Color online) Level diagram of the Raman processes generating the synthetic spin-orbit coupling. The Raman laser beams are counterpropagating along $\vec{e}_{x}$ with frequencies $\omega_i$ and $\omega_i+\Delta\omega_i$ ($i=1,2$). The two-photon detuning is $\delta$, and $\epsilon$ is the quadratic Zeeman shift~\cite{spielman2009}.}
\label{Level}
\end{figure}
\begin{figure}[tbp]
\includegraphics[width=8cm]{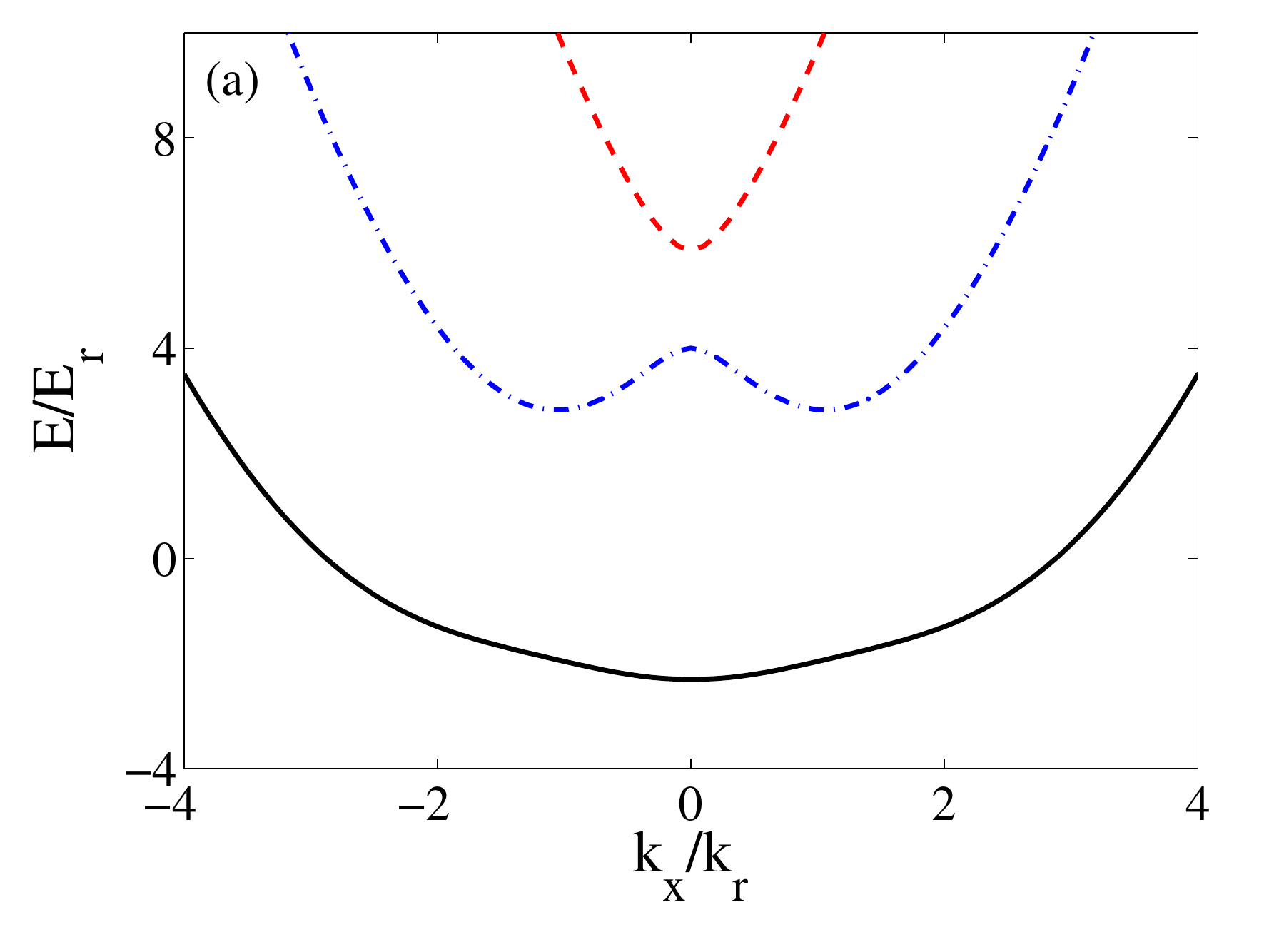}
\includegraphics[width=8cm]{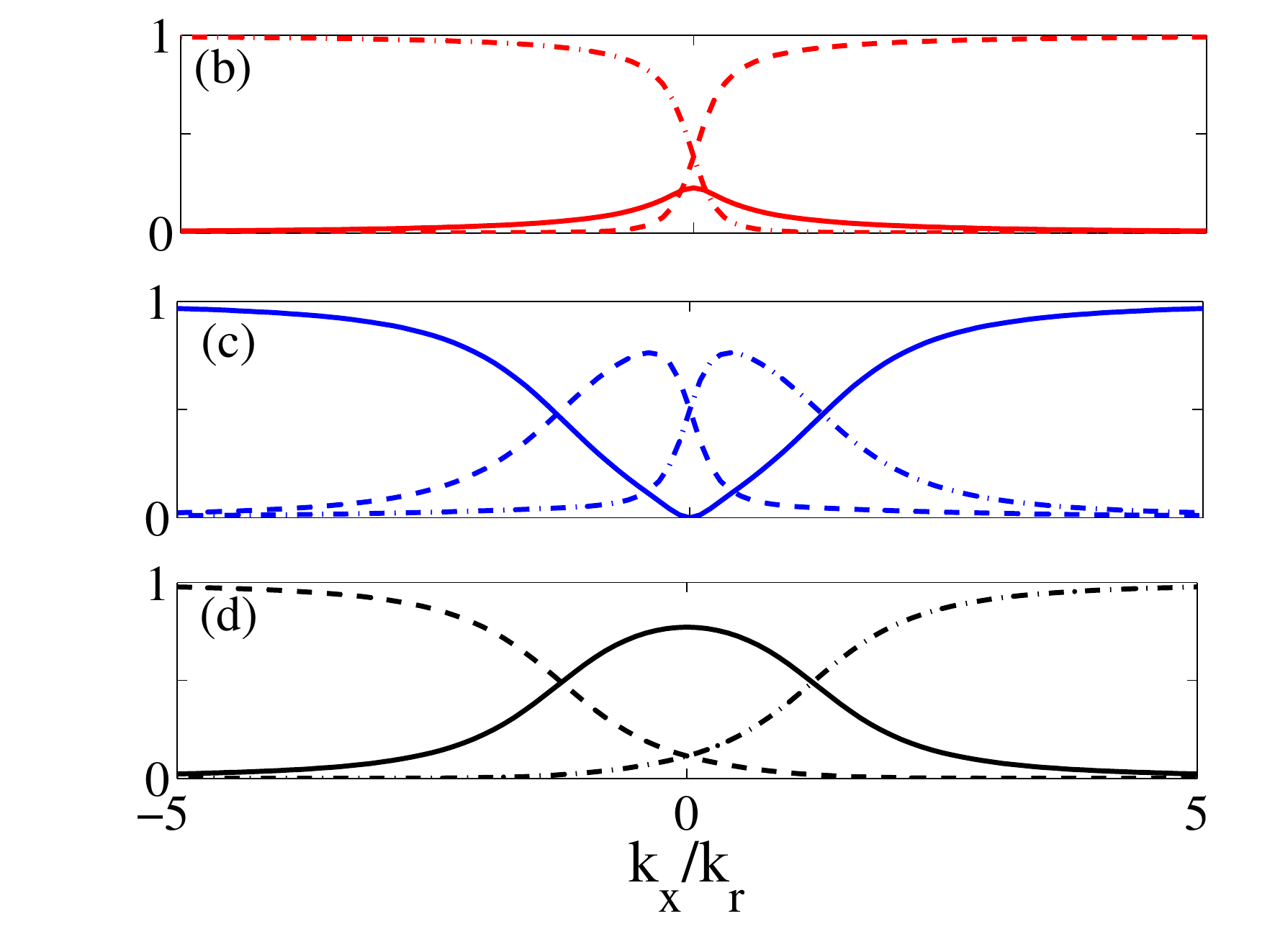}
\caption{(Color online) (a) Single-particle energy spectra along the $k_y=0$ axis in momentum space for a homogeneous gas. The red dashed curve denotes $E_1$, the blue dashed-dotted curve is $E_2$, and the black solid one represents $E_3$. (b)-(d) The momentum distributions of the hyperfine states in different helicity branches. The dashed curves denote $|a_{i}|^2$, the solid curves are $|b_{i}|^2$, and the dashed-dotted ones represent $|c_{i}|^2$. Here, $i=1-3$ for (b)-(d), respectively. The parameters are $\hbar\epsilon=0.44E_{r}$, $\hbar\delta=0$, and $h=2.425E_{r}$~\cite{spielman2009}.}
\label{single}
\end{figure}

We consider a two dimensional three-component Fermi gas where the three atomic hyperfine states are labeled as $|\pm 1\rangle$ and $|0\rangle$. The hyperfine states are coupled by Raman lasers as illustrated in Fig.~\ref{Level}. Under an appropriate rotating frame, the single-particle Hamiltonian can be written in momentum space as~\cite{zhaireview2012,spielmanreview,spielman2009}
\begin{eqnarray}\label{H0}
&&\textsl{H}_{0}(\vec{k})=\nonumber\\
        &&\left(
          \begin{array}{ccc}
            \frac{\hbar^{2}(\vec{k}+2k_{r}\vec{e}_{x})^{2}}{2m}-\hbar\delta & h & 0 \\
            h & \frac{\hbar^{2}k^{2}}{2m}-\hbar\epsilon & h \\
            0 & h & \frac{\hbar^{2}(\vec{k}-2k_{r}\vec{e}_{x})^{2}}{2m}+\hbar\delta \\
          \end{array}
        \right),\nonumber\\
\end{eqnarray}
where $\delta$ is the two-photon detuning of the Raman process, $\epsilon$ accounts for a small quadratic Zeeman shift, and $h=\hbar\Omega_{R}/2$ denotes the strength of the effective Zeeman field, which is proportional to the Rabi frequency of the Raman process $\Omega_{R}$. We take the recoil energy $E_{r}=\hbar^{2}k^{2}_{r}/(2m)$ and the corresponding wave vector $k_{r}$ as the units of energy and wave vector, respectively, with $m$ as the atomic mass. In the rest of the paper, we will only consider the case where $\delta=0$.
The Hamiltonian $\textsl{H}_{0}(\vec{k})$ can be diagonalized as
\begin{eqnarray}
\textsl{H}_{0}(\vec{k})\left[
          \begin{array}{c}
            a_i(\vec{k}) \\
            b_i(\vec{k}) \\
            c_i(\vec{k}) \\
          \end{array}
        \right]=
          E_i(\vec{k})\left[
          \begin{array}{c}
            a_i(\vec{k}) \\
            b_i(\vec{k}) \\
            c_i (\vec{k})\\
          \end{array}
          \right],
\end{eqnarray}
where the eigenvalues $E_{i}(\vec{k})$ ($i=1-3$) are the single-particle dispersion spectra for the helicity branches, and $\psi_{i}(\vec{k})=(a_{i},b_{i},c_{i})^{T}$ are the corresponding eigenvectors. Importantly, the coefficients $a_i(\vec{k})$, $b_i(\vec{k})$, and $c_i(\vec{k})$ are related to the weight of hyperfine states in the corresponding helicity branches.

In Fig.~\ref{single}, we plot the single-particle dispersion spectra as well as the momentum distributions of the hyperfine states in the helicity branches along the $k_y=0$ axis. Apparently, the hyperfine-state superpositions in the helicity branches are momentum dependent [Figs.~\ref{single}(b)-\ref{single}(d)]. A critical observation is that although the momentum distribution of the hyperfine state $|0\rangle$ is symmetric with respect to $k_x=0$ for all three helicity branches, such a symmetry is absent for the cases of states $|1\rangle$ and $|-1\rangle$. Instead, the momentum distributions of $|1\rangle$ and $|-1\rangle$ are symmetric with respect to each other such that $|a_i(k_x)|^2=|c_i(-k_x)|^2$. The presence or absence of this inversion symmetry has crucial effects on pairing physics. In fact, in the weak-coupling limit, pairing tends to occur between two fermions residing on their corresponding Fermi surfaces in the absence of SOC. With SOC mixing up the hyperfine spins into the helicity branches, should we turn on a small attractive interaction between hyperfine states $|1\rangle$ and $|-1\rangle$, the pairing state would have zero center-of-mass momentum. On the other hand, if we turn on a small attractive interaction between states $|1\rangle$ and $|0\rangle$ (or $|-1\rangle$ and $|0\rangle$), the  pairing state would have a finite center-of-mass momentum. This is similar to the FF pairing mechanism in Ref.~\cite{yi2014PRL} where the interplay of SOC and spin-selective interactions plays the key role.

\begin{figure*}[tbp]
\includegraphics[width=5.6cm]{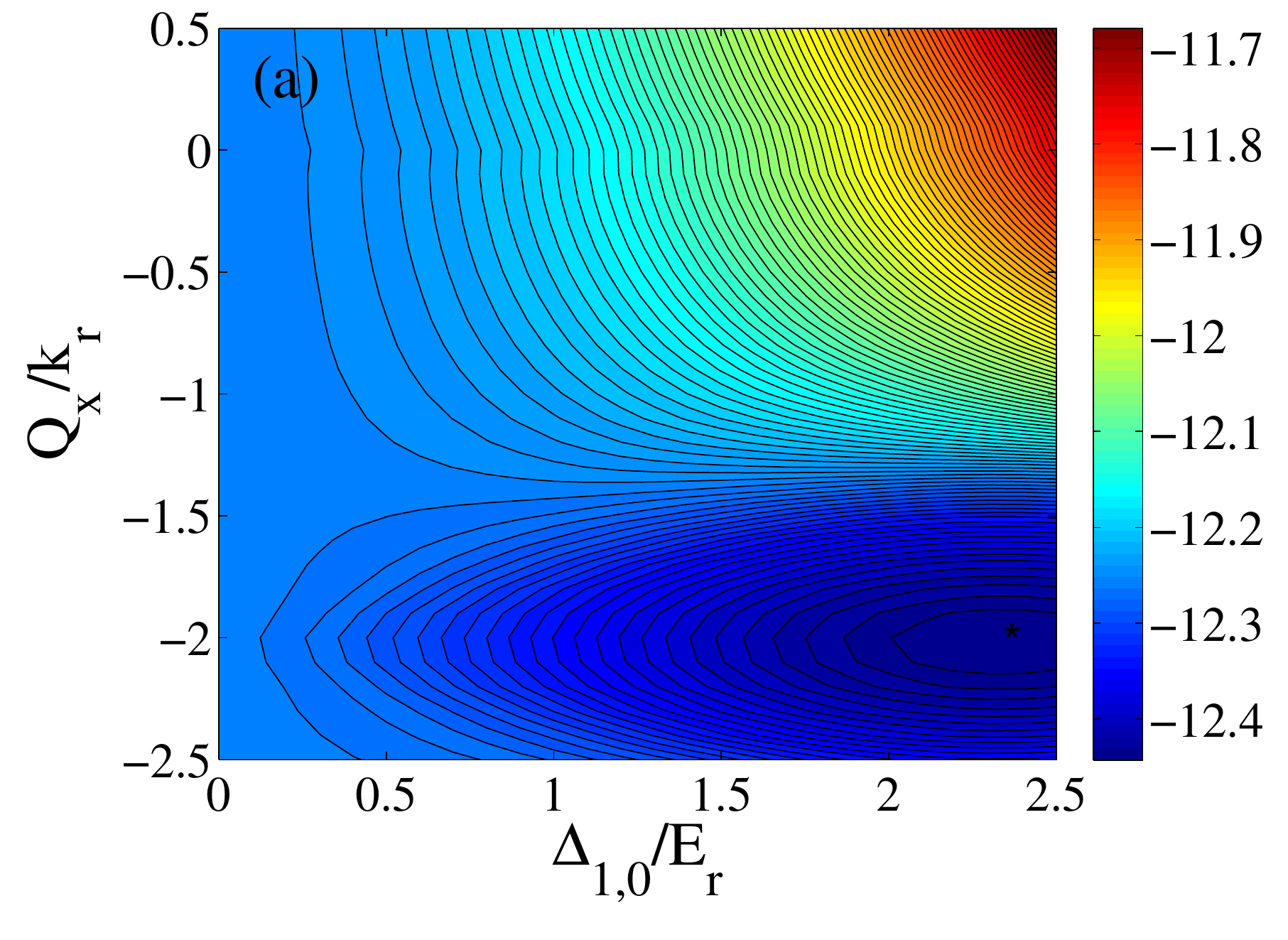}
\includegraphics[width=5.6cm]{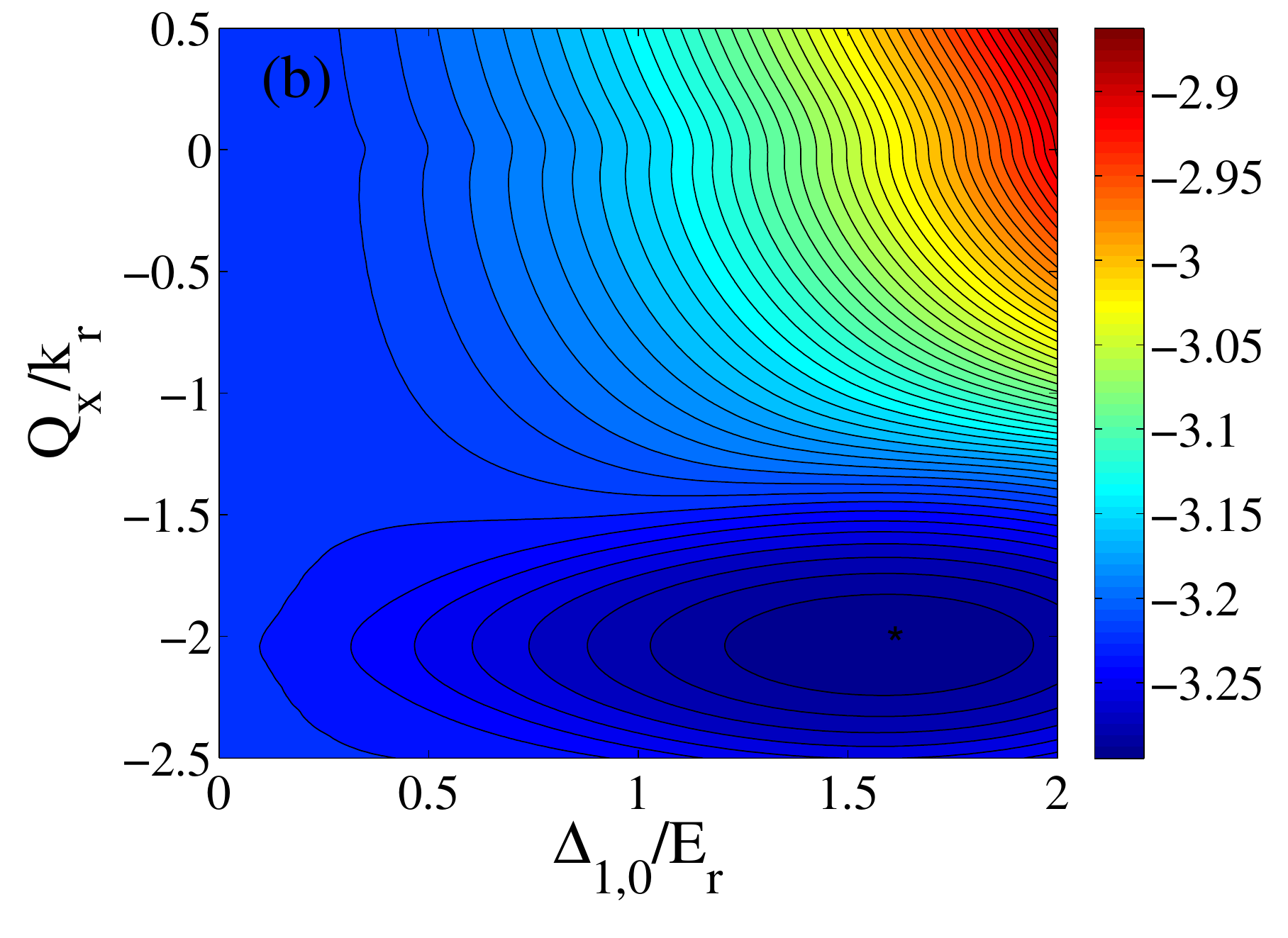}
\includegraphics[width=5.6cm]{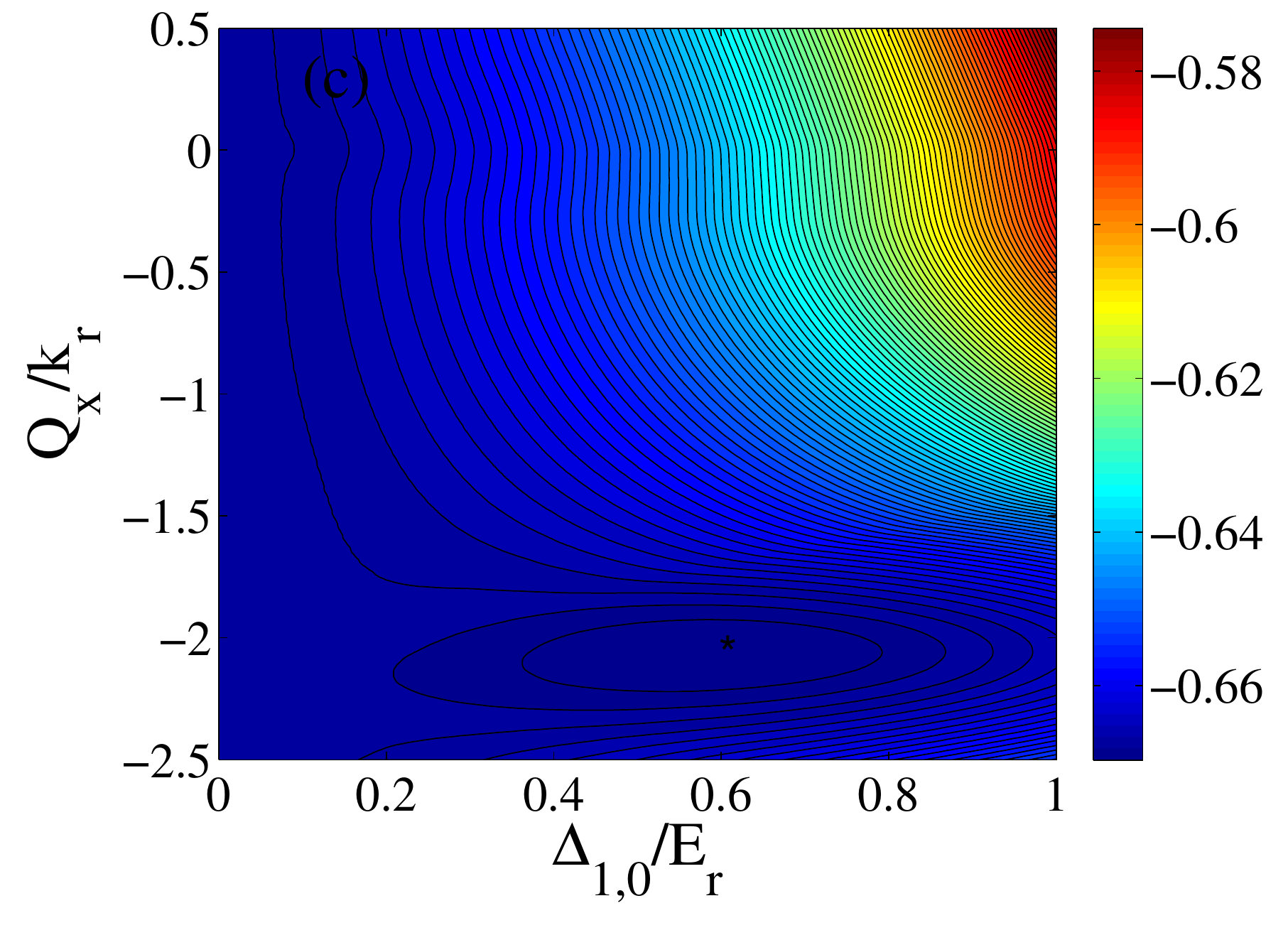}
  \caption{(Color online) Contour plots of the thermodynamic potential landscape in the $\Delta_{1,0}$-$Q_x$ plane for (a) $\mu=10E_r$ and $E_{b1}=0.3E_r$; (b) $\mu=5E_r$ and $E_{b1}=0.3E_r$; (c) $\mu=2E_r$ and $E_{b1}=0.3E_r$. The global minima are located at (a) ($\Delta_{1,0}/E_r\sim2.3410$, $Q_x/k_r\sim-2.0184$); (b) ($\Delta_{1,0}/E_r\sim1.5917$, $Q_x/k_r\sim-2.0347$); (c) ($\Delta_{1,0}/E_r\sim0.5970$, $Q_x/k_r\sim-2.0666$). Other parameters are chosen as $\hbar\epsilon=0.2E_{r}$, $\hbar\delta=0$, and $h=E_{r}$.}
\label{OmegaFF}
\end{figure*}

\section{Mean-field treatment of pairing states}\label{3}

With the understanding of the single-particle dispersion, we now study the many-body pairing physics using the standard mean-field formalism.
Now the complete Hamiltonian can be written as
\begin{eqnarray}
H=\sum_{\vec{k}} \Phi^{\dag}(\vec{k})[H_{0}(\vec{k})-\mu]\Phi(\vec{k}) +\textsl{H}_{\rm I},\end{eqnarray}
where $\Phi^{\dagger}(\vec{k})=(c^{\dagger}_{\vec{k},1},c^{\dagger}_{\vec{k},0},c^{\dagger}_{\vec{k},-1})$
is the creation operator for the three hyperfine states, $c_{\vec{k},i}$ is an annihilation operator of a Fermi atom with wave vector
$\vec{k}$ and pseudospin $i=1,0,-1$ describing three atomic hyperfine states.
$c^{\dag}_{\vec{k},i}$ is the corresponding creation operator, and $\mu$ is the chemical potential.
Here, the most general form of the interaction Hamiltonian $H_I$ is
\begin{eqnarray}\label{HIk}
\textsl{H}_{\rm I}&&=\frac{g_{1,0}}{V}\sum_{\vec{Q},\vec{k},\vec{k'}}c^{\dagger}_{\vec{k},1}c^{\dagger}_{\vec{Q}-\vec{k},0}c_{\vec{k'},0}c_{\vec{Q}-\vec{k'},1}\nonumber\\
&&~~+\frac{g_{1,-1}}{V}\sum_{\vec{Q},\vec{k},\vec{k'}}c^{\dagger}_{\vec{k},1}c^{\dagger}_{\vec{Q}-\vec{k},-1}c_{\vec{k'},-1}c_{\vec{Q}-\vec{k'},1}\nonumber\\
&&~~+\frac{g_{0,-1}}{V}\sum_{\vec{Q},\vec{k},\vec{k'}}c^{\dagger}_{\vec{k},0}c^{\dagger}_{\vec{Q}-\vec{k},-1}c_{\vec{k'},-1}c_{\vec{Q}-\vec{k'},0}.
\end{eqnarray}
Here, $\vec{Q}$ is the center-of-mass momentum of the pairing fermions, $V$ is the quantization volume in two dimensions, and $g_{i,j}$ ($i,j=\pm 1,0$) is the bare interaction rate between different hyperfine states, which can be renormalized following the standard procedure in two dimensions \cite{2drenorm,cuirenorm,pzhangrenorm},
\begin{equation}\label{g2D}
\frac{1}{g_{i,j}}=-\frac{1}{V}\sum_{\vec{k}}\frac{1}{\hbar^{2}k^{2}/m+E_{b,i,j}}.
\end{equation}
Here, $E_{b,i,j}$ is the two-body bound state energy between states $|i\rangle$ and $|j\rangle$. The two-body bound state energy can be related to the corresponding three-dimensional $s$-wave scattering length,
which can be tuned experimentally via the Feshbach resonance technique.

Under the mean-field approximation, the interaction term becomes
\begin{eqnarray}
\textsl{H}_{\rm I}&&\thickapprox\sum_{\vec{k}}\big(\Delta_{\vec{Q},1,0}c^{\dagger}_{\vec{k},1} c^{\dagger}_{\vec{Q}-\vec{k},0}+\Delta^{*}_{\vec{Q},1,0}c_{\vec{Q}-\vec{k},0} c_{\vec{k},1}\big)\nonumber\\
&&~~+\sum_{\vec{k}}\big(\Delta_{\vec{Q},1,-1}c^{\dagger}_{\vec{k},1} c^{\dagger}_{\vec{Q}-\vec{k},-1}+\Delta^{*}_{\vec{Q},1,-1}c_{\vec{Q}-\vec{k},-1} c_{\vec{k},1}\big)\nonumber\\
&&~~+\sum_{\vec{k}}\big(\Delta_{\vec{Q},0,-1}c^{\dagger}_{\vec{k},0} c^{\dagger}_{\vec{Q}-\vec{k},-1}+\Delta^{*}_{\vec{Q},0,-1}c_{\vec{Q}-\vec{k},-1} c_{\vec{k},0}\big)\nonumber\\
&&~~-\frac{V|\Delta_{\vec{Q},1,0}|^2}{g_{1,0}}-\frac{V|\Delta_{\vec{Q},1,-1}|^2}{g_{1,-1}}-\frac{V|\Delta_{\vec{Q},0,-1}|^2}{g_{0,-1}},
\end{eqnarray} where the superfluid order parameter is taken as
\begin{eqnarray}\label{Delta}
\Delta_{\vec{Q},i,j}=\frac{g_{i,j}}{V}\sum_{\vec{k}}\langle c_{\vec{Q}-\vec{k},j}c_{\vec{k},i}\rangle,~~~(i,j=1,0,-1).
\end{eqnarray}
Therefore, in the hyperfine-spin basis $\Psi_{\vec{Q}}(\vec{k})=(c_{\vec{k},1},c^{\dagger}_{
\vec{Q}-\vec{k},1},c_{\vec{k},0},c^{\dagger}_{\vec{Q}-\vec{k},0},c_{\vec{k},-1},c^{\dagger}_{
\vec{Q}-\vec{k},-1})^{T}$, the effective mean-field Hamiltonian can be rewritten as
\begin{eqnarray}\label{Hm}
H_{m}&&=\frac{1}{2}\sum_{\vec{k}}\Psi_{\vec{Q}}^{\dagger }(\vec{k})M_{\vec{k}}\Psi_{\vec{Q}}(\vec{k})\nonumber\\
&&~~+\sum_{\vec{k}}\left(\frac{3}{2}\xi_{\vec{Q}-\vec{k}}+4E_{r}-\frac{1}{2}\hbar\epsilon\right)\nonumber\\
&&~~-V\left(\frac{|\Delta_{\vec{Q},1,0}|^{2}}{g_{1,0}}+\frac{|\Delta_{\vec{Q},1,-1}|^{2}}{g_{1,-1}}+\frac{|\Delta_{\vec{Q},0,-1}|^{2}}{g_{0,-1}}\right).\nonumber\\
\end{eqnarray}
The matrix takes the form of
\begin{widetext}
\begin{eqnarray}\label{M}
&&M_{\vec{k}}=\left(
\begin{array}{cccccc}
\xi_{\vec{k}+2k_{r}\vec{e}_{x}}-\hbar\delta & 0 & h & \Delta_{\vec{Q},1,0} & 0 & \Delta_{\vec{Q},1,-1} \\
0 & -(\xi_{\vec{Q}-\vec{k}+2k_{r}\vec{e}_{x}}-\hbar\delta) & -\Delta^{*}_{\vec{Q},1,0} & -h & -\Delta^{*}_{\vec{Q},1,-1} & 0 \\
h & -\Delta_{\vec{Q},1,0} & \xi_{\vec{k}}-\hbar\epsilon & 0 & h & \Delta_{\vec{Q},0,-1} \\
\Delta^{*}_{\vec{Q},1,0} & -h  & 0 & -(\xi_{\vec{Q}-\vec{k}}-\hbar\epsilon) & -\Delta^{*}_{\vec{Q},0,-1} & -h \\
0 & -\Delta_{\vec{Q},1,-1} & h & -\Delta_{\vec{Q},0,-1} & \xi_{\vec{k}-2k_{r}\vec{e}_{x}}+\hbar\delta & 0 \\
\Delta^{*}_{\vec{Q},1,-1} & 0 & \Delta^{*}_{\vec{Q},0,-1} & -h & 0 & -(\xi_{\vec{Q}-\vec{k}-2k_{r}\vec{e}_{x}}+\hbar\delta)
\end{array}
\right),
\end{eqnarray}
\end{widetext}
with $\xi_{\vec{k}}=\hbar^2k^2/(2m)-\mu$.

The zero-temperature thermodynamic potential can then be derived
\begin{eqnarray}\label{omega}
\Omega
&&=-\left.k_{B}T\ln{\text{Tr}{e^{-H_{m}/(k_{B}T)}}}\right|_{T\rightarrow 0}\nonumber\\
&&=\frac{1}{2}\sum_{\vec{k},j=1,2,3,4,5,6}E_{\vec{k},j}\Theta(-E_{\vec{k},j})\nonumber\\
&&~~+\sum_{\vec{k}}\left(\frac{3}{2}\xi_{\vec{Q}-\vec{k}}+4E_{r}-\frac{1}{2}\hbar\epsilon\right)\nonumber\\
&&~~-V\left(\frac{|\Delta_{\vec{Q},1,0}|^{2}}{g_{1,0}}+\frac{|\Delta_{\vec{Q},1,-1}|^{2}}{g_{1,-1}}+\frac{|\Delta_{\vec{Q},0,-1}|^{2}}{g_{0,-1}}\right),\nonumber\\
\end{eqnarray}
where $\Theta(x)$ is the Heaviside step function, $\text{Tr}$ denotes the trace over both the momentum and the spin degrees of freedom, $T$ represents the temperature, and $k_{B}$ is the Boltzmann constant.

In the following, we first consider the simple case where inter-atomic interactions are limited between two spin species. Due to the symmetry of the setup at $\delta=0$, we only need to study two different cases among all three possible combinations of interactions: the case with interaction between $|1\rangle$ and $|0\rangle$ and the case with interaction between $|1\rangle$ and $|-1\rangle$.

\section{Pairing states with interaction existing between two spin species}\label{4}

We first study the case where interaction only presents between states $|1\rangle$ and $|0\rangle$ so that $\Delta_{\vec{Q},1,-1}=\Delta_{\vec{Q},0,-1}=0$. We also define $\Delta_{\vec{Q},1,0} \equiv \Delta_{1,0}$ and $E_{b1} \equiv E_{b,1,0}$ to simplify notations. Under these conditions, we calculate the thermodynamic potential in Eq. (\ref{omega}) for any given $Q$ by numerically diagonalizing the matrix Eq. (\ref{M}). For the parameters that we have studied, the local minimum in the thermodynamic potential always occurs with $\vec{Q}=Q_x \vec{e}_{x}$, where $\vec{e}_{x}$ is the unit vector along the direction of the SOC. In Figs.~\ref{OmegaFF}(a)-\ref{OmegaFF}(c), we show the typical contour plots of the thermodynamic potential on the plane of $\Delta_{1,0}$-$Q_x$ for a given chemical potential $\mu$ and binding energy $E_{b1}$. From these figures, we can see clearly that in general there exists only one local minimum, which corresponds to the ground state of the system. This is consistent with our previous analysis that under SOC-induced asymmetric hyperfine spin distribution and spin-selective interaction between $|1\rangle$ and $|0\rangle$, the pairing state acquires a nonzero center-of-mass momentum, i.e., it is an FF state.

\begin{figure}[tbp]
\includegraphics[width=4.5cm]{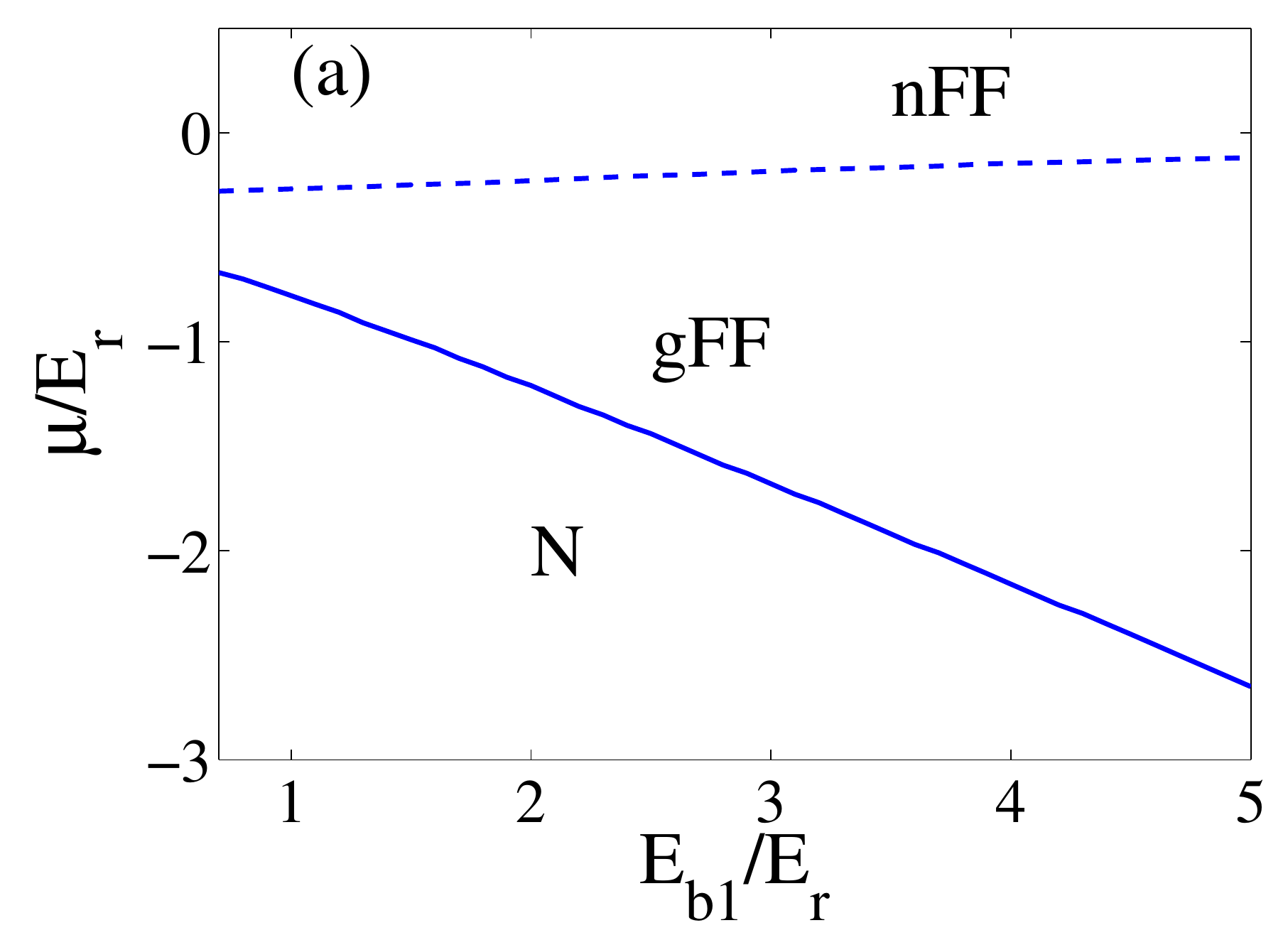}\includegraphics[width=4.5cm]{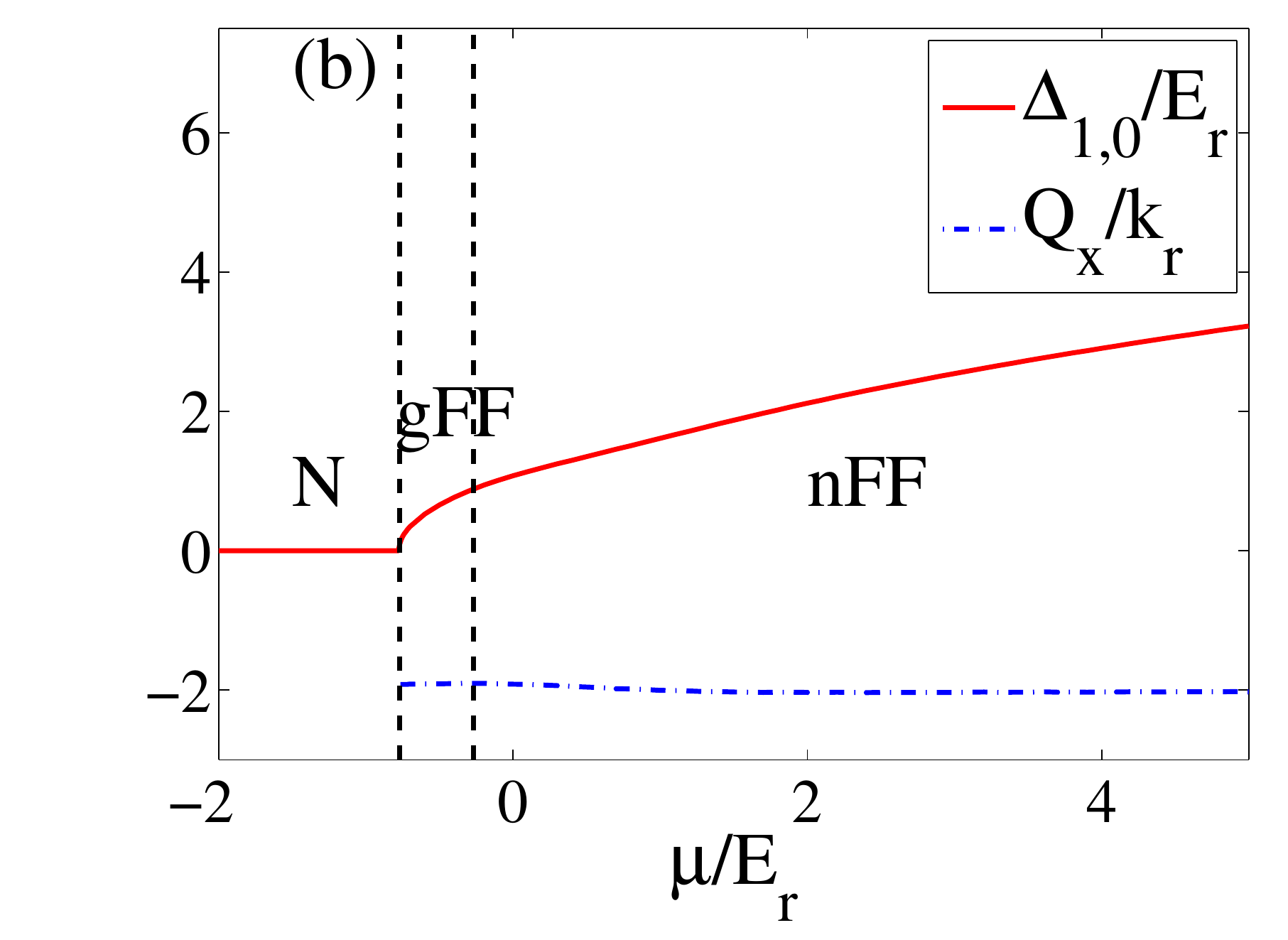}\\
\includegraphics[width=4.5cm]{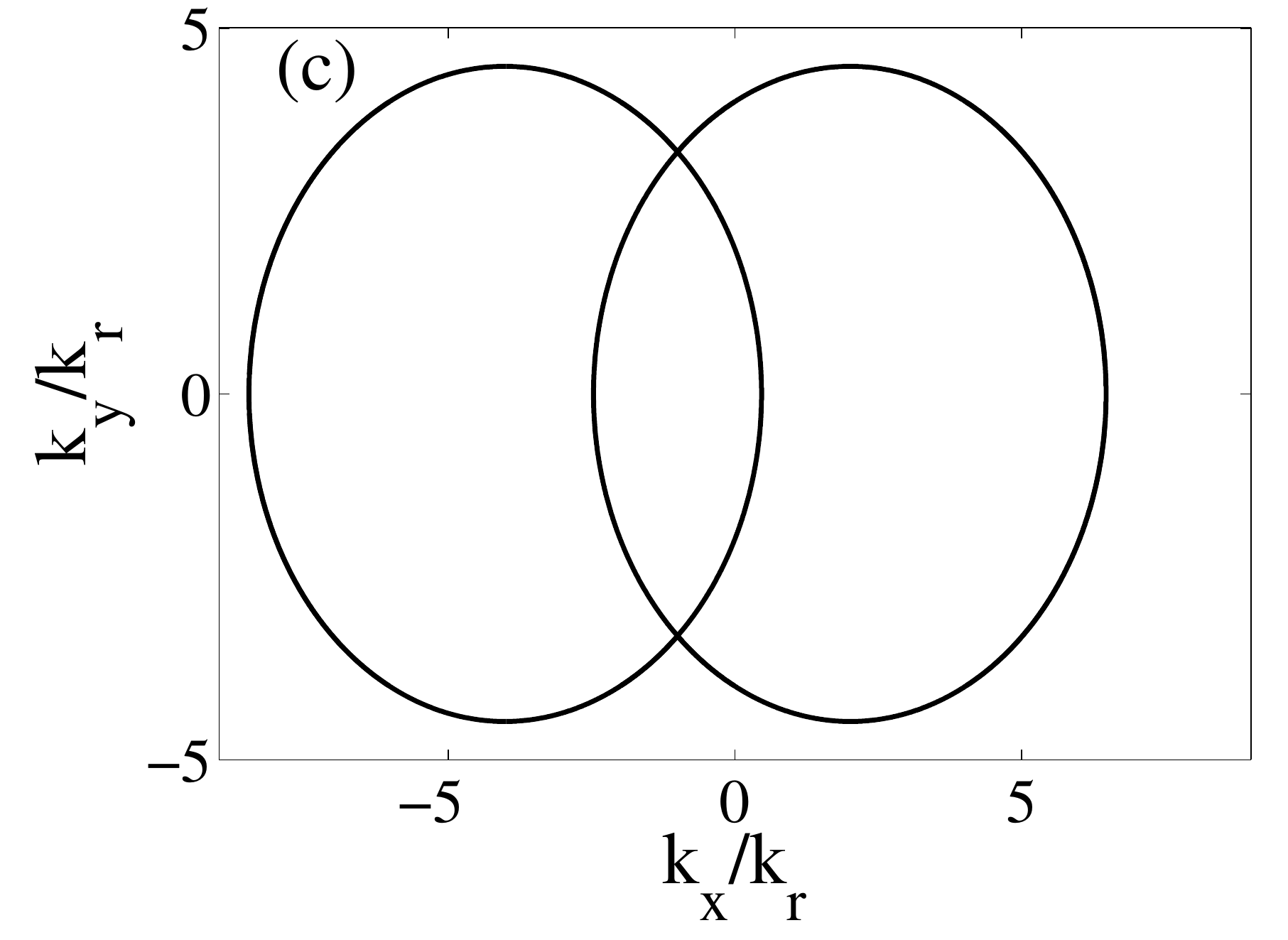}\includegraphics[width=4.5cm]{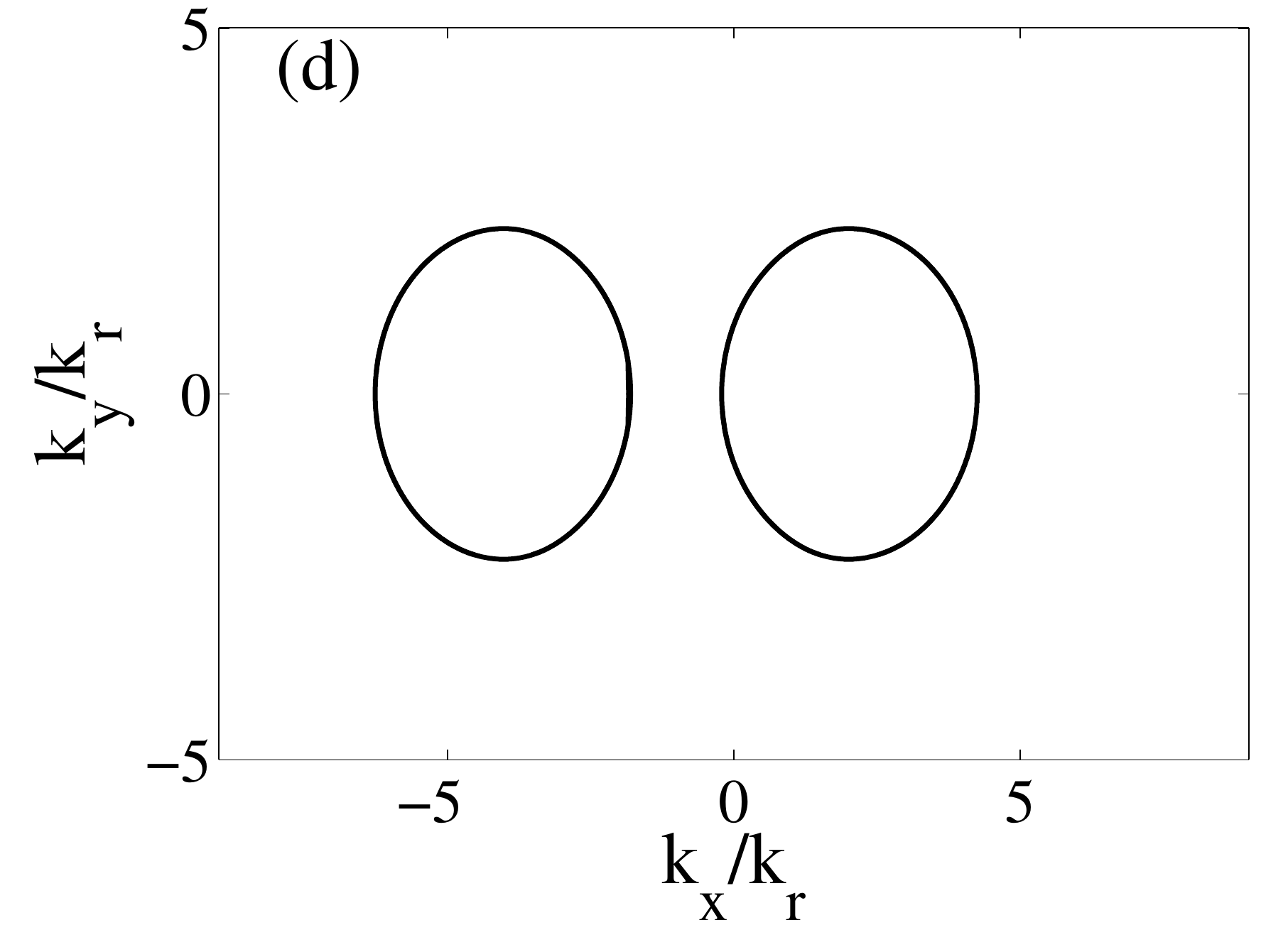}
  \caption{(Color online) (a) Phase diagram on the $\mu$-$E_{b1}$ plane for $\hbar\epsilon=0.2E_{r}$, $\hbar\delta=0$,
and $h=E_{r}$. The blue solid curve represents the phase boundary of the continuous phase from the FF state to the normal state,
and it marks the $|\Delta_{1,0}|=10^{-3}E_r$ threshold.
The blue dashed curve denotes the phase boundary between the fully gapped FF state and the gapless nodal FF state.
(b) Typical ground state superfluid order parameter $\Delta_{1,0}$ and center momentum $Q_x$ of the pairing fermions versus $\mu$ for $\hbar\epsilon=0.2E_{r}$, $\hbar\delta=0$, $h=E_{r}$ and $E_{b1}=E_r$.
Typical contours of gapless points in momentum space for the gapless nodal FF state with
(c) $\mu=20E_r$, $E_{b1}=2E_r$ (nFF); (d) $\mu=5E_r$, $E_{b1}=2E_r$ (nFF).}
\label{FF}
\end{figure}

These observations allow us to minimize the thermodynamic potential with respect to the order parameter $\Delta_{1,0}$ and the center-of-mass momentum $Q=Q_x$ to find the ground state of the system. In Fig.~\ref{FF}(a), we map out the phase diagram on the $\mu$-$E_{b1}$ plane for $\hbar\epsilon=0.2E_{r}$, $\hbar\delta=0$, and $h=E_{r}$. Apparently, a continuous phase boundary exists between the FF state, which is characterized by a finite $\Delta_{1,0}$ and a finite $Q_x$, and the normal state ($N$), which is characterized by a vanishing order parameter. Furthermore, judging from the minimum excitation gap, both a fully gapped FF state (gFF) and a gapless nodal FF (nFF) state exist on the phase diagram, which are separated by a continuous phase boundary. In Fig.~\ref{FF}(b), we show how the ground-state order parameter $\Delta_{1,0}$ and the center-of-mass momentum $Q_x$ evolve with the chemical potential $\mu$ with fixed $h=E_{r}$ and $E_{b1}=E_r$. In Figs.~\ref{FF}(c) and \ref{FF}(d), we also show the typical gapless contours of the nFF state in momentum space.

For the case where the only interaction in the system is between states $|1\rangle$ and $|-1\rangle$, we may set $\Delta_{\vec{Q},1,0}=\Delta_{\vec{Q},0,-1}=0$. By diagonalizing the effective Hamiltonian and minimizing the thermodynamic potential as we have performed previously, we find that the ground state of the system is either a BCS pairing state with $Q=0$ or a normal state with a vanishing order parameter. This is also consistent with our previous analysis in the weakly interacting limit that attractive interactions between states $|1\rangle$ and $|-1\rangle$ would lead to BCS pairing due to the symmetry in the momentum distribution of these states in the helicity branches.

\begin{figure}[tbp]
\centering
\includegraphics[width=8cm]{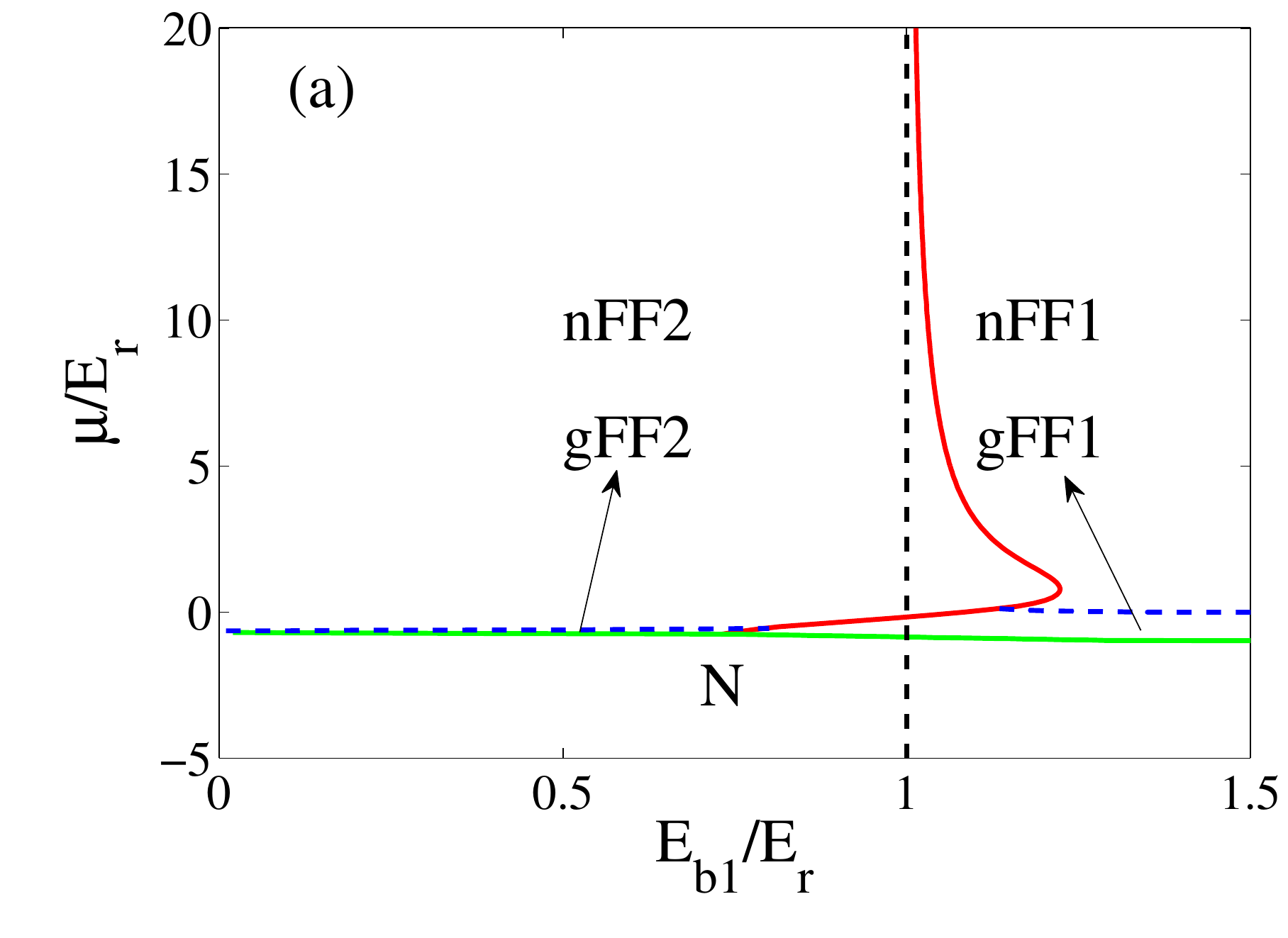}
\includegraphics[width=8cm]{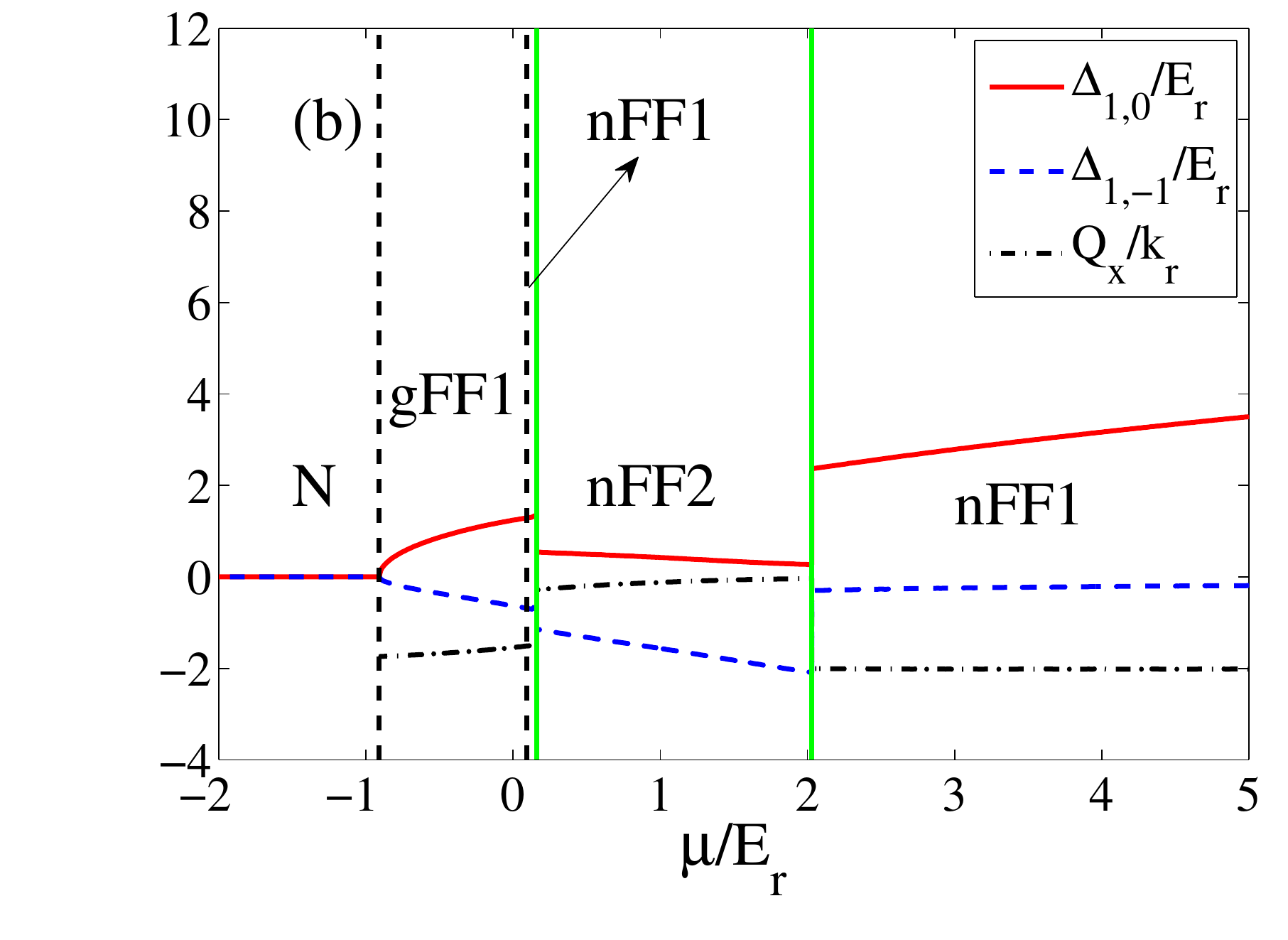}
  \caption{(Color online) (a) Phase diagram on the $\mu$-$E_{b1}$ plane for $\hbar\epsilon=0.2E_{r}$, $\hbar\delta=0$, $h=E_{r}$, and $E_{b2}=E_{r}$. The red solid curve is the first-order boundary, whereas the green (light gray) solid curve represents the phase boundary of the continuous phase from the FF state to the normal state, and the blue dashed curves denote the phase boundaries between the fully gapped FF state and the gapless nodal FF state. The black dashed line denotes $E_{b1}=E_{r}$, and it is the asymptote of the first-order boundary. (b) Typical ground state superfluid order parameters $\Delta_{1,0}$, $\Delta_{1,-1}$ and center momentum $Q_x$ of the pairing fermions versus $\mu$ for $\hbar\epsilon=0.2E_{r}$, $\hbar\delta=0$, $h=E_{r}$, $E_{b2}=E_{r}$, and $E_{b1}=1.15E_r$. The red solid curve is $\Delta_{1,0}$, the blue dashed curve represents $\Delta_{1,-1}$, and the black dashed-dotted curve denotes $Q_x$.}
\label{NFF}
\end{figure}

\section{Pairing states with multiple interactions among various spin combinations}\label{5}

In this section, we consider the case where attractive interactions are not only present between
states $|1\rangle$ and $|0\rangle$, but also exist between states $|1\rangle$ and $|-1\rangle$.
Hence, we can set $\Delta_{\vec{Q},0,-1}=0$, $\Delta_{\vec{Q},1,-1}=\Delta_{1,-1}$, and $E_{b2}=E_{b,1,-1}$ for convenience.

\begin{figure*}[tbp]
\includegraphics[width=4.3cm]{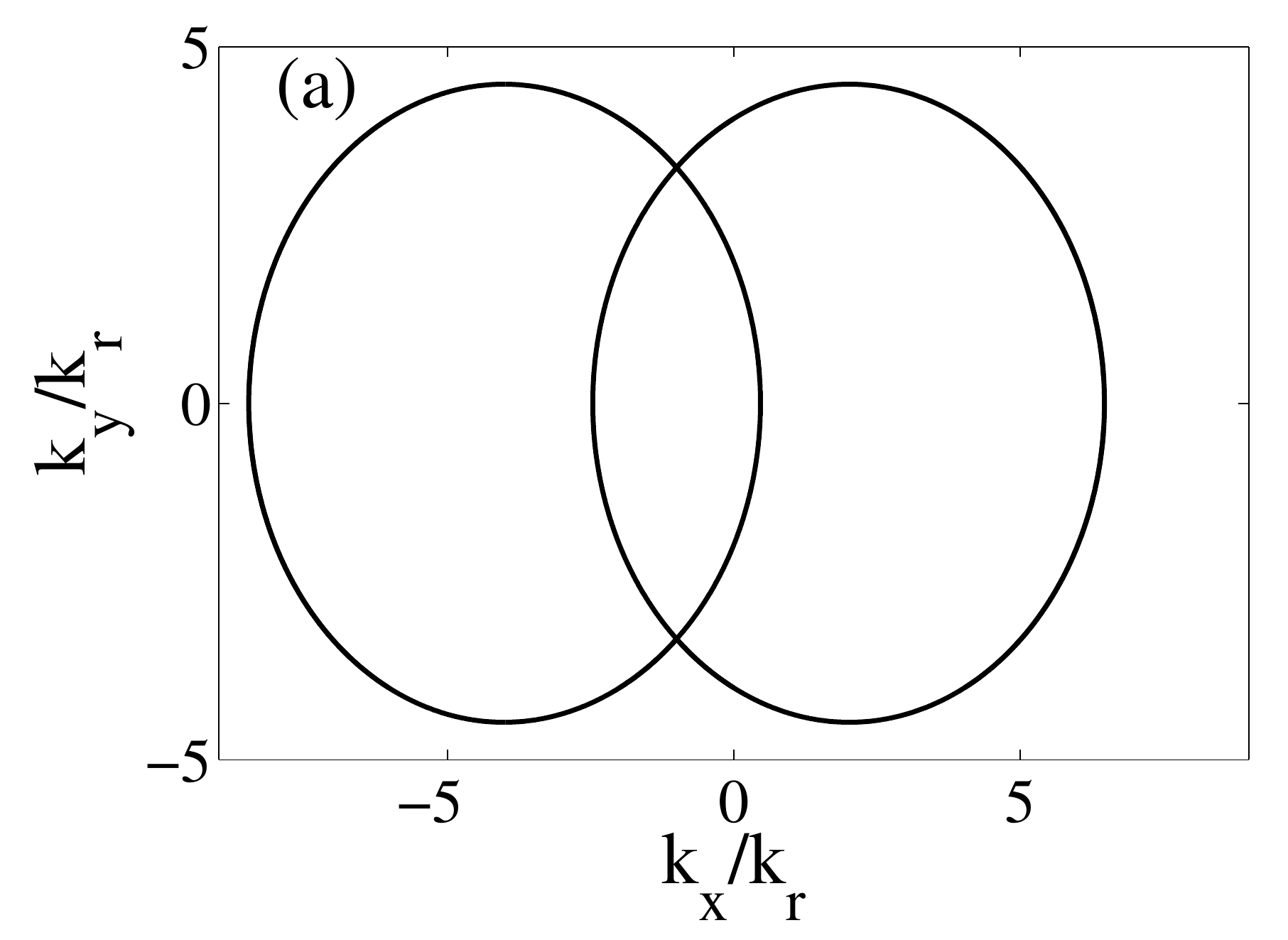}
\includegraphics[width=4.3cm]{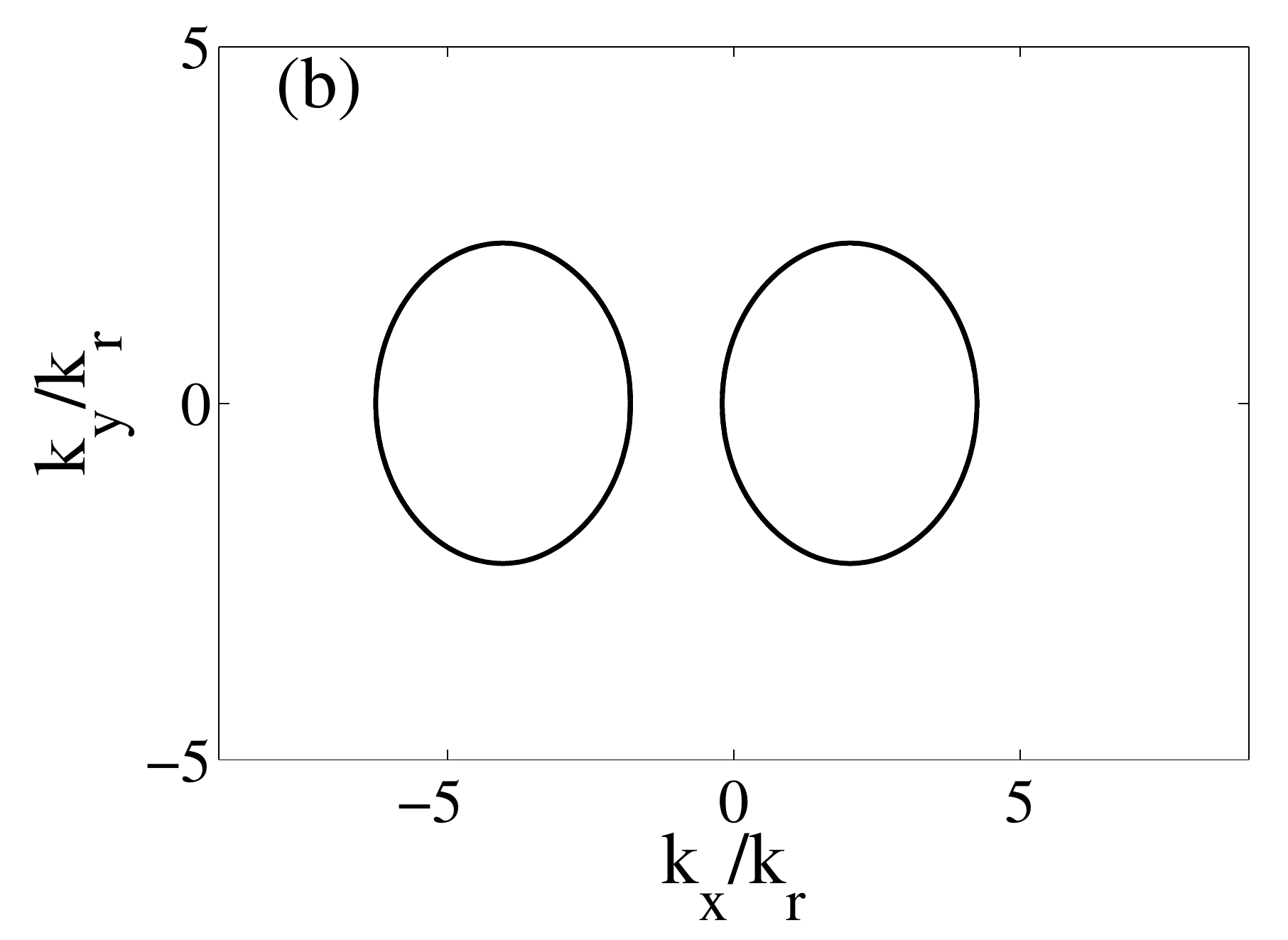}
\includegraphics[width=4.3cm]{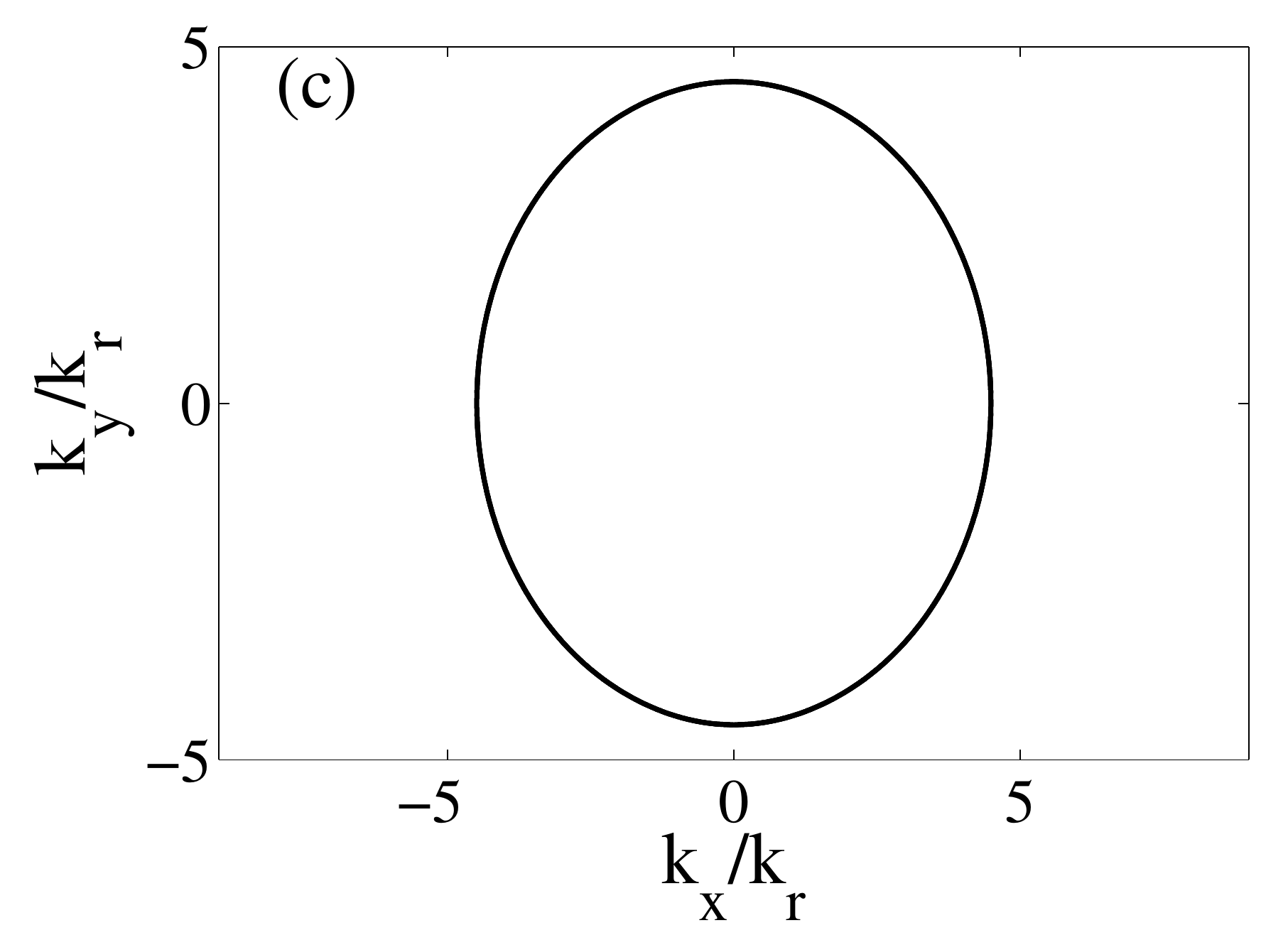}
\includegraphics[width=4.3cm]{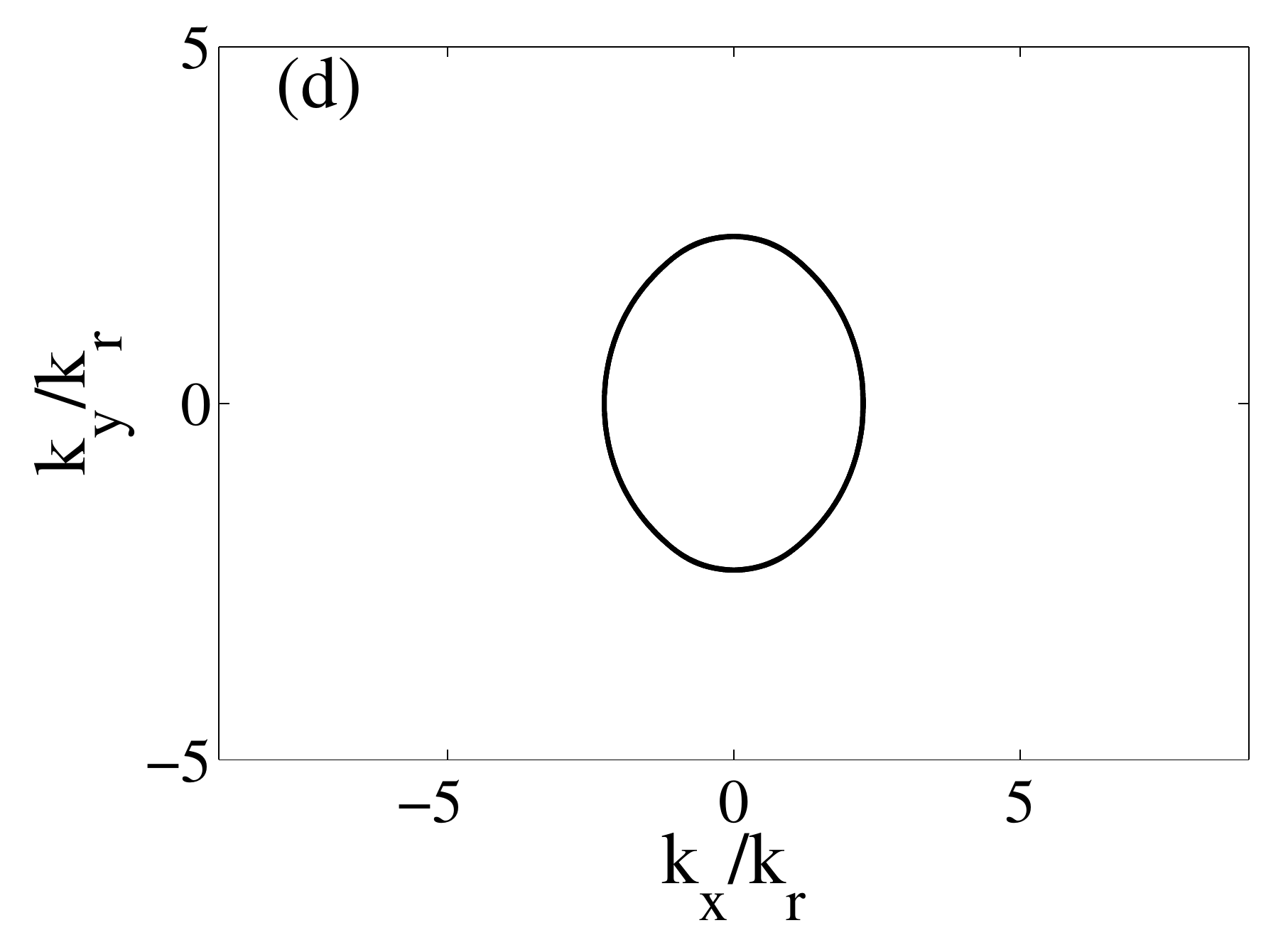}\\
\includegraphics[width=4.3cm]{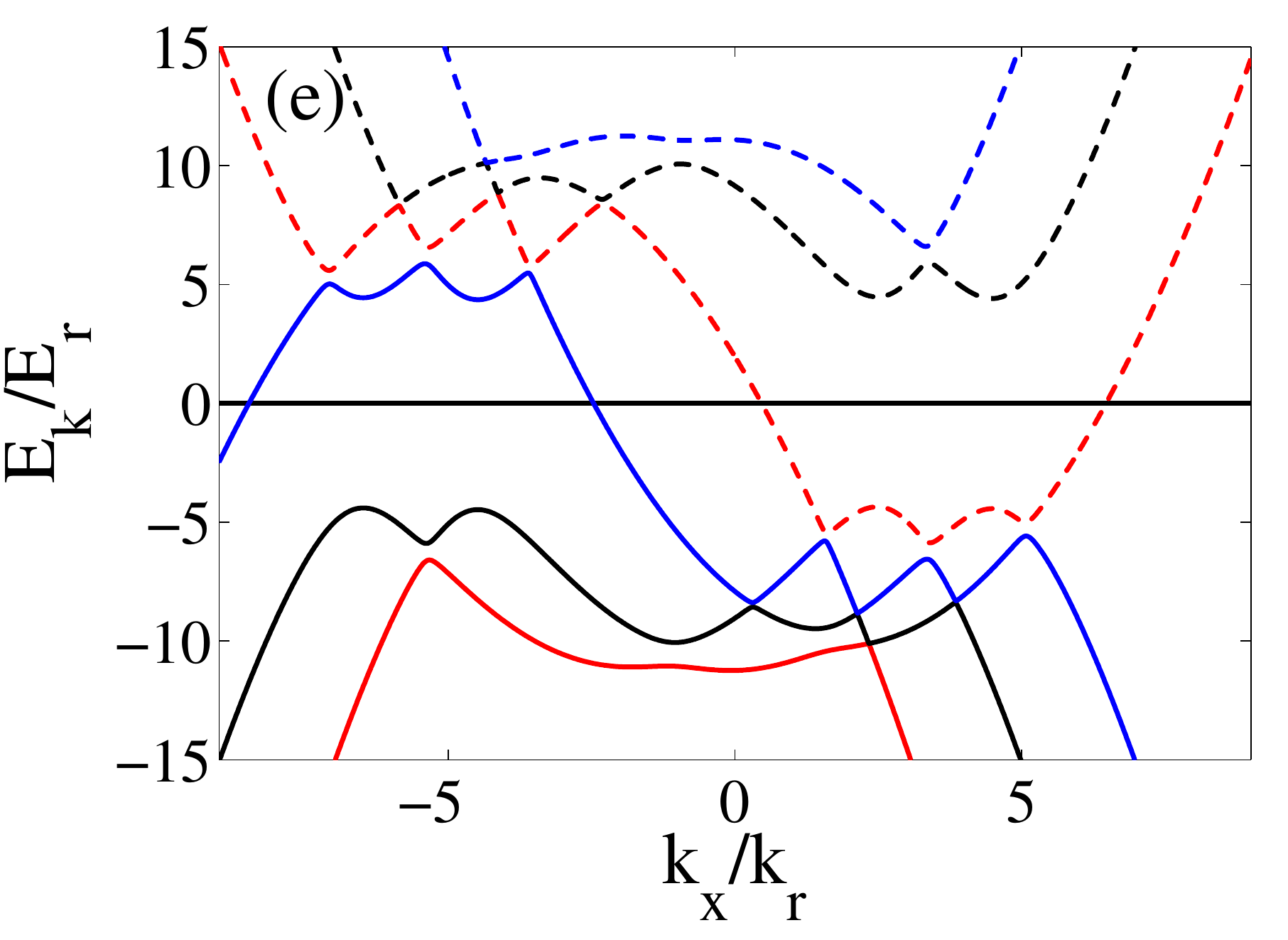}
\includegraphics[width=4.3cm]{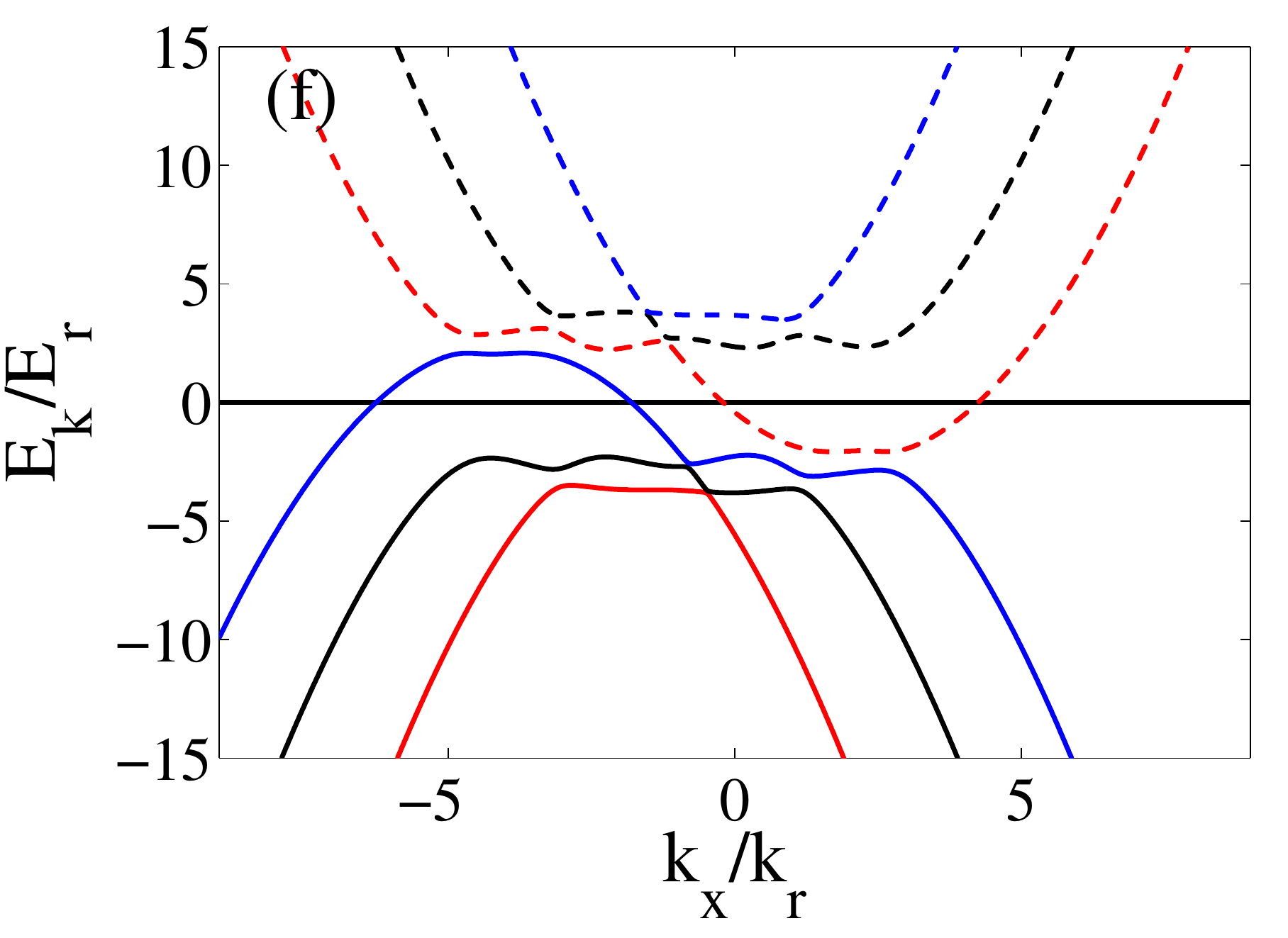}
\includegraphics[width=4.3cm]{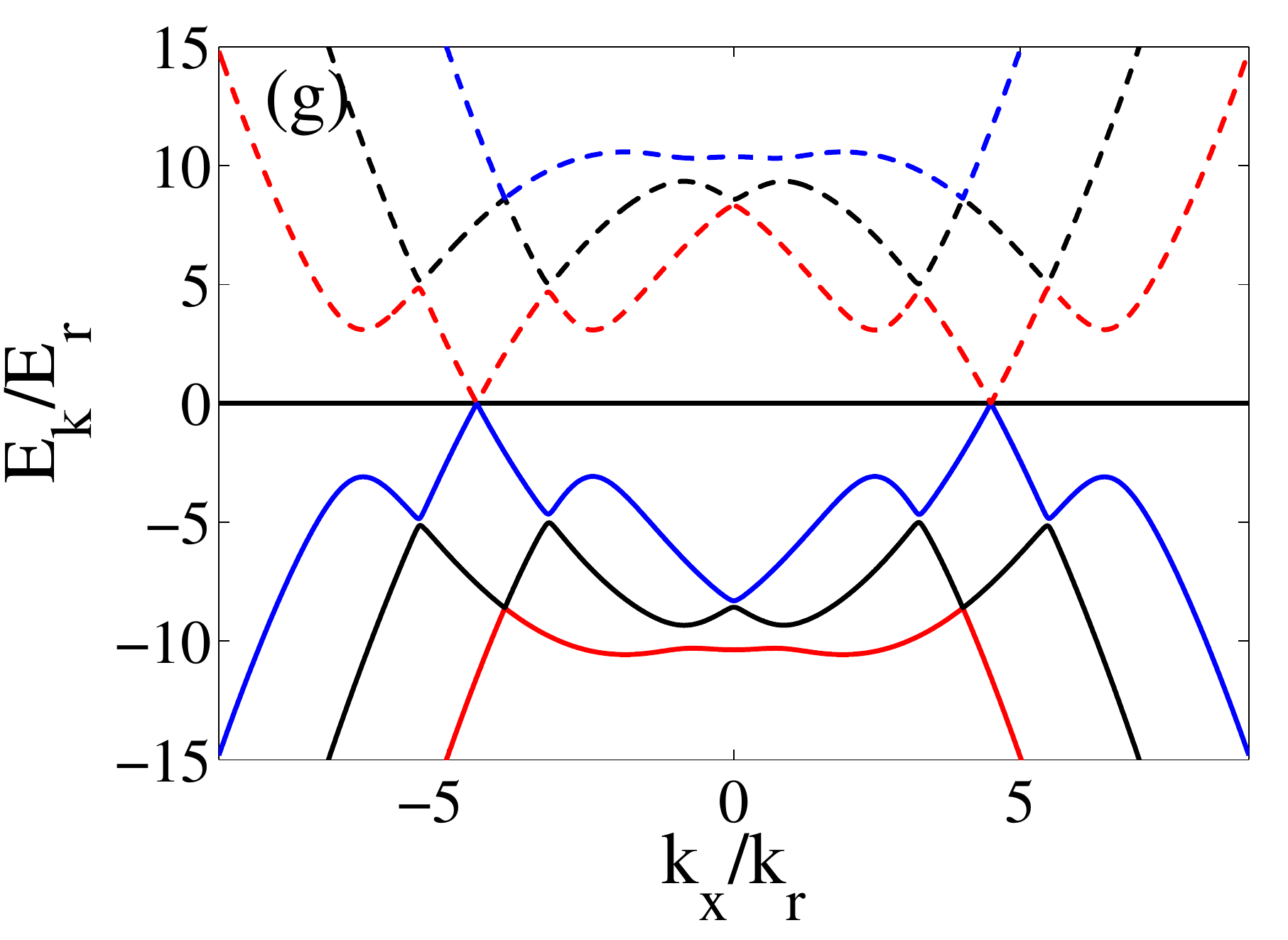}
\includegraphics[width=4.3cm]{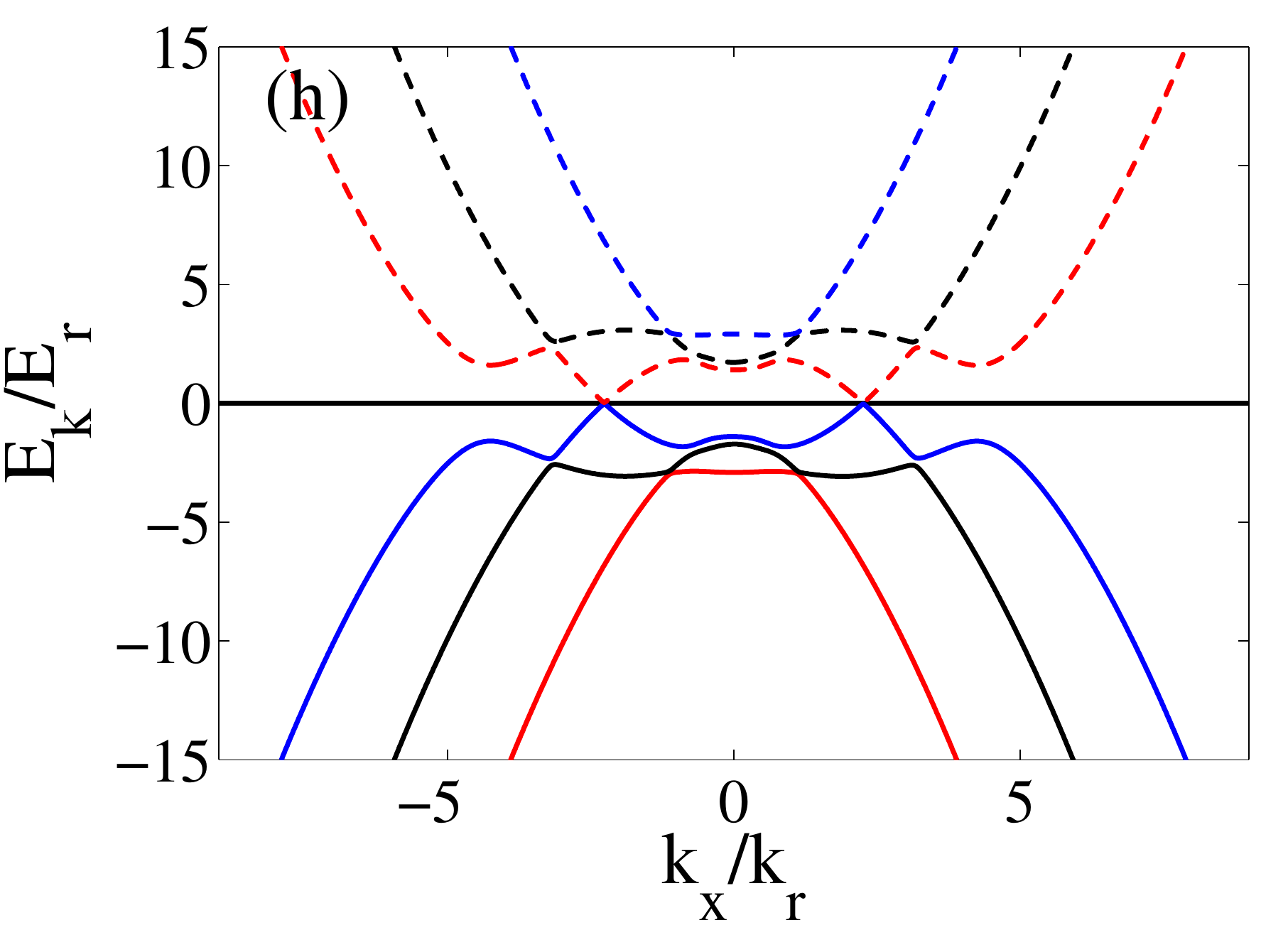}
\caption{(Color online) Contours of gapless points in momentum space for different novel gapless nodal FF states with
(a) $\mu=20E_r$, $E_{b1}=2E_r$ (nFF1); (b) $\mu=5E_r$, $E_{b1}=2E_r$ (nFF1);
(c) $\mu=20E_r$, $E_{b1}=0.2E_r$ (nFF2); (d) $\mu=5E_r$, $E_{b1}=0.2E_r$ (nFF2). (e)-(h) Quasiparticle (quasihole) dispersion spectra along the $k_y=0$ axis for the novel gapless nodal FF states that correspond to panels (a)-(d), respectively. Other parameters are chosen as $\hbar\epsilon=0.2E_{r}$, $\hbar\delta=0$, $h=E_{r}$, and $E_{b2}=E_{r}$.}
\label{gaplessNFF}
\end{figure*}

\begin{figure*}[tbp]
\includegraphics[width=4.3cm]{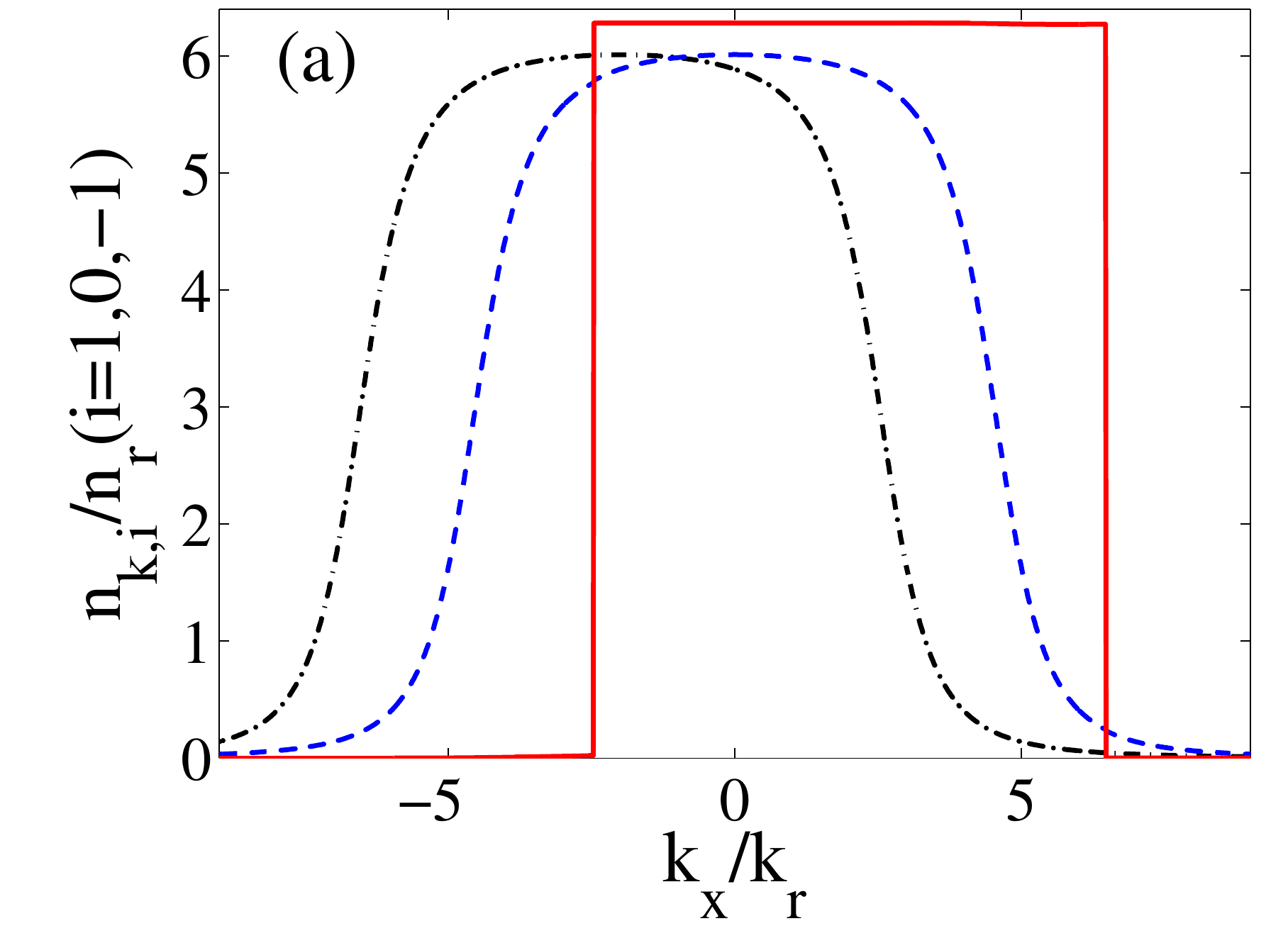}
\includegraphics[width=4.3cm]{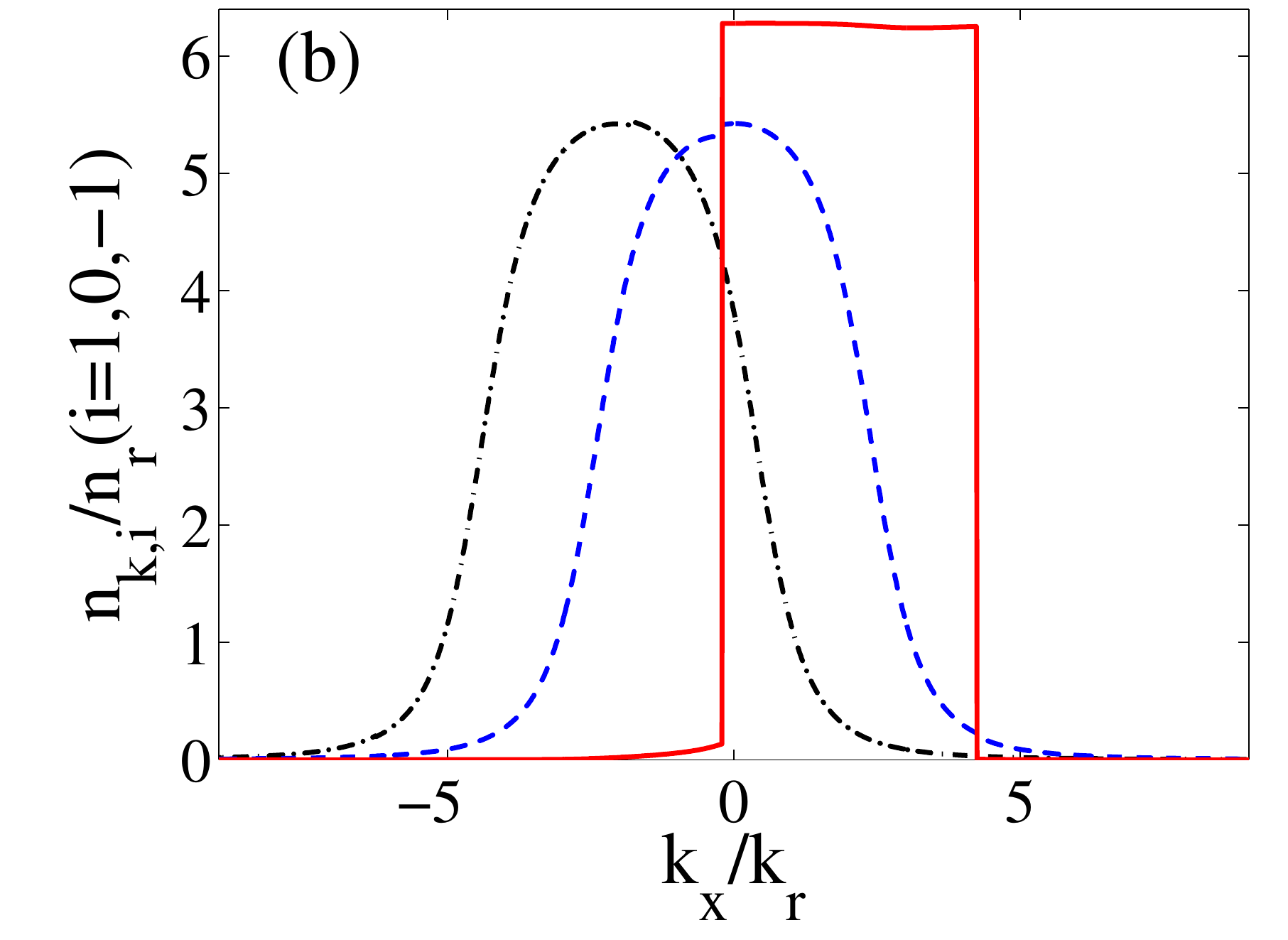}
\includegraphics[width=4.3cm]{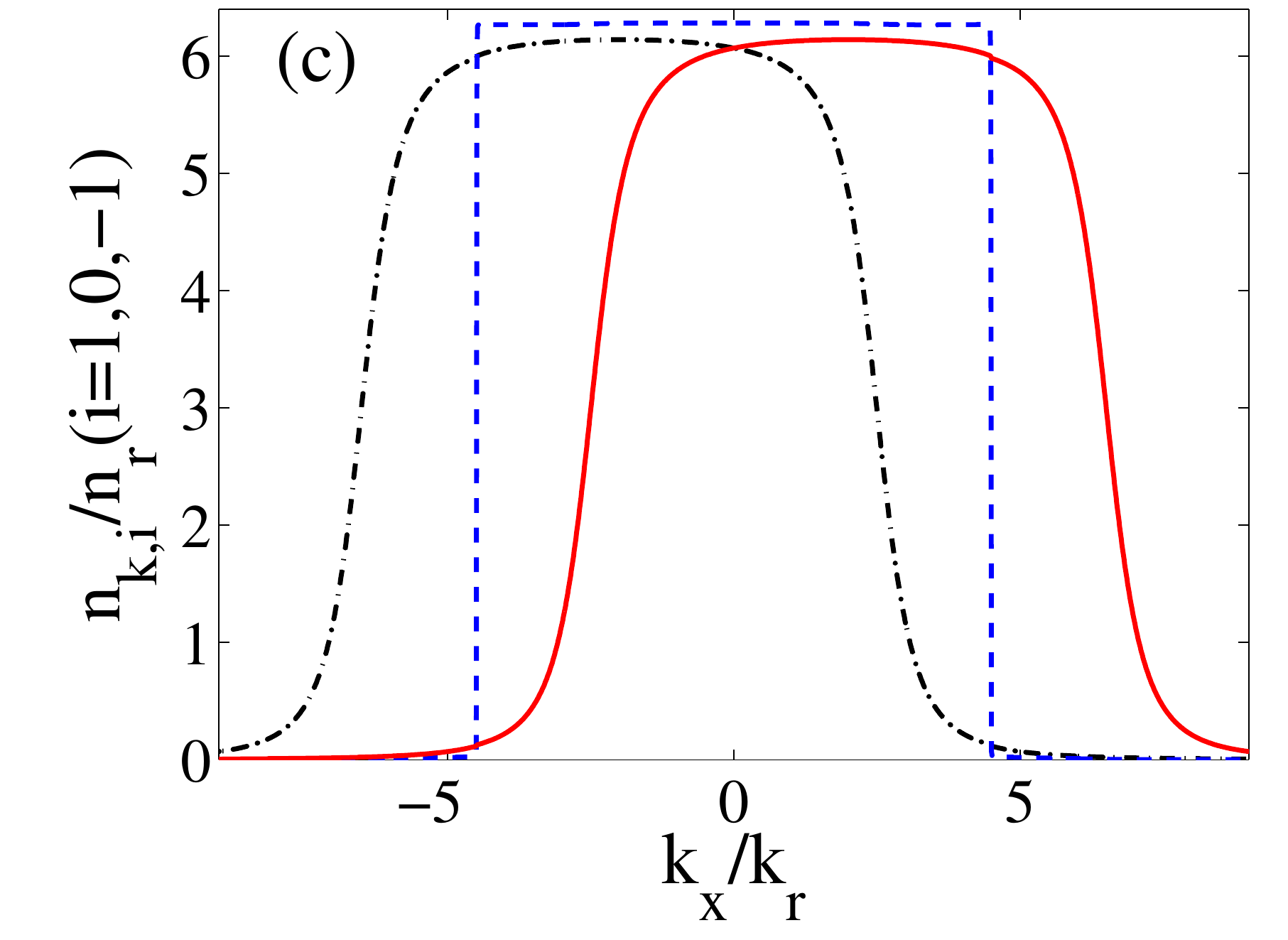}
\includegraphics[width=4.3cm]{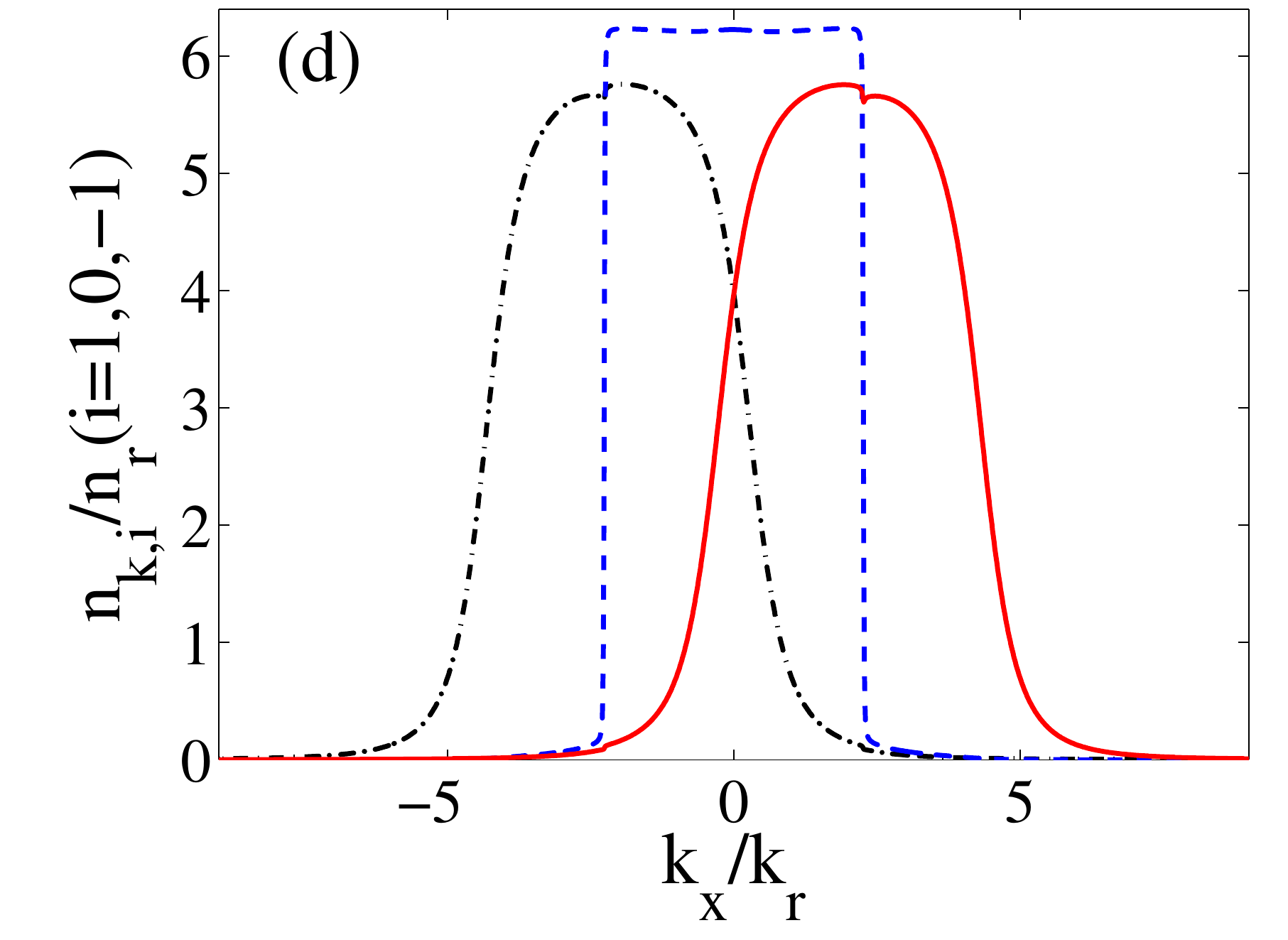}
\caption{(Color online) Number density distributions in momentum space along the $k_y=0$ axis for different novel gapless nodal FF states with
(a) $\mu=20E_r$, $E_{b1}=2E_r$ (nFF1); (b) $\mu=5E_r$, $E_{b1}=2E_r$ (nFF1); (c) $\mu=20E_r$, $E_{b1}=0.2E_r$ (nFF2); (d) $\mu=5E_r$, $E_{b1}=0.2E_r$ (nFF2). The black dashed-dotted curve is $n_{k,1}/n_r$, the blue dashed curve represents $n_{k,0}/n_r$, and the red solid curve denotes $n_{k,-1}/n_r$. Other parameters are chosen as $\hbar\epsilon=0.2E_{r}$, $\hbar\delta=0$, $h=E_{r}$, and $E_{b2}=E_{r}$. The unit of density is defined through $n_{r}=k^{2}_{r}/(2\pi)$.}
\label{densityNFF}
\end{figure*}

As before, we numerically minimize the thermodynamic potential to look for the ground state of the system. In Fig.~\ref{NFF}(a), we show the typical phase diagram on the $\mu$-$E_{b1}$ plane for $\hbar\epsilon=0.2E_{r}$, $\hbar\delta=0$, $h=E_{r}$, and $E_{b2}=E_{r}$. It can be clearly seen from Fig.~\ref{NFF}(a) that there are four distinct FF phases on the phase diagram, including two nodal (or gapless) (labeled as nFF1 and nFF2) and two fully gapped (gFF1 and gFF2) ones. Whereas the gapped FF states are separated from the nodal FF states by continuous phase boundaries [blue dashed lines in Fig.~\ref{NFF}(a)], different nodal FF states or different gapped FF states are separated by a first-order phase boundary [red solid line in Fig.~\ref{NFF}(a)]. In this phase diagram, we also identify a normal state ($N$) by numerically setting a small threshold of $|\Delta|=10^{-3}E_r$.

By adjusting the binding energy $E_{b1}$, the system can be tuned across different phases over a wide range of $\mu$. In Fig.~\ref{NFF}(b), we show how the ground-state order parameters $\Delta_{1,0}$, $\Delta_{1,-1}$ and the center-of-mass momentum $Q_x$ evolve with
the chemical potential $\mu$ with fixed $h=E_{r}$, $E_{b2}=E_{r}$, and $E_{b1}=1.15E_r$. For this typical parameter set, the system can successively go through multiple phase transitions by increasing $\mu$. In the local-density approximation where the effect of a global trapping potential is taken into account by the spatial variation in chemical potential, this is the order of phases that one would observe starting from a trap edge to its center.

To further characterize the properties of these FF states, we demonstrate in Fig.~\ref{gaplessNFF} the typical gapless contours and dispersion spectra in momentum space. For the gapless contours, we find that there are two closed gapless rings in nFF1, and the two rings are both symmetric about the $k_y=0$ axis [see Figs.~\ref{gaplessNFF}(a) and \ref{gaplessNFF}(b)]. By decreasing the chemical potential $\mu$, the two rings become smaller in size, and gradually separate from each other. The gapless contours of nFF2 are shown in Figs.~\ref{gaplessNFF}(c) and \ref{gaplessNFF}(d). Different from the two gapless rings in nFF1, there is only one gapless ring in nFF2 with given parameters. This gapless ring is symmetric with respect to the origin. As the chemical potential decreases, the gapless ring becomes smaller. We also show in Figs.~\ref{gaplessNFF}(e)-\ref{gaplessNFF}(h) the quasiparticle and quasihole dispersion spectra of nFF1 and nFF2 along the $k_y=0$ axis. These are consistent with the corresponding gapless contours in Figs.~\ref{gaplessNFF}(a)-\ref{gaplessNFF}(d). In principle, one may probe these dispersion spectra experimentally using momentum-resolved radio-frequency spectroscopy.

The interesting features of the gapless contours and the dispersion spectra also leave signatures in the particle-number distribution in momentum space, which may be probed more directly via the time-of-flight measurement. We show in Fig.~\ref{densityNFF} the number density distributions in momentum space along the $k_y=0$ axis for different nodal FF states with the same parameters as in Fig.~\ref{gaplessNFF}. For the nFF1 case with $E_{b1}=2E_r$, it can be clearly seen from Figs.~\ref{densityNFF}(a) and \ref{densityNFF}(b) that abrupt changes are present in the momentum-space density profiles, particularly for the $|-1\rangle$ state. Comparing to the corresponding results of gapless contours, one can find that the discontinuous features are consistent with the right half of the gapless contours. For the nFF2 phase as shown in Figs.~\ref{densityNFF}(c) and \ref{densityNFF}(d), similar discontinuities can be found in the momentum distribution of the $|0\rangle$ state, which are consistent with the corresponding structure of gapless contours.

\section{Summary}\label{6}

We have studied the properties of the SOC-induced FF pairing states in a two-dimensional three-component Fermi gas with the recently realized synthetic SOC at zero temperature. The FF state here is the result of SOC-induced asymmetric momentum distribution of hyperfine states and the spin-selective interaction, both of which are experimentally achievable. To illustrate this, we investigate in detail the impact of different combinations of spin-selective interactions on the properties of the pairing states of the system. Interestingly, the interplay of SOC and spin-selective interaction can give rise to a novel three-component FF state in which every two of the three components form an FF pairing state with a common center-of-mass momentum. We study in detail the stability region of the FF states, the dispersion spectra of quasiparticle and quasihole excitations, the gapless contours and the number distributions in momentum space, and discuss possible experimental detection schemes based on our results. As both the synthetic SOC and the spin-selective interactions have been realized experimentally, our paper has interesting implications for future experiments on SOC-induced exotic superfluidity. In particular, we stress that even at temperatures above the superfluid transition temperature, this spin-selective pairing mechanism would assist the emergence of two-body bound states under appropriate interaction parameters. These two-body bound states should then acquire a finite center-of-mass momentum depending on the details of the interaction. As a consequence, the exotic pairing physics discussed above can be verified via a radio-frequency spectroscopy analysis of two-body bound states in the normal phase.

\section*{Acknowledgements}
We thank Jian-Song Pan and Zeng-Qiang Yu for helpful discussions. This work is supported by NFRP (Grants No. 2011CB921200 and Grants No. 2011CBA00200), NNSF (Grants No. 60921091), NSFC (Grants No. 11274009, No. 11434001, No. 11404106, and No. 11374283). F.Q. acknowledges support from the Guidance Project of Education Department of Hubei Province under Grant No. B2014024, the Teaching Reform Research Project of Hubei Polytechnic University under Grant No. 2014C16, and the Scientific Research Foundation of Hubei Polytechnic University under Grant No. 14xjz04R. W.Y. acknowledges support from the ``Strategic Priority Research Program(B)'' of the Chinese Academy of Sciences, Grant No. XDB01030200. W.Z. thanks the Research Funds of Renmin University of China (Grants No. 10XNL016) for support.

\end{CJK*}
\end{document}